\documentclass[aps, prd, twocolumn,a4paper]{revtex4-1}
\usepackage{ulem}
\usepackage{color}
\usepackage{float}
\usepackage{appendix}
\usepackage{graphicx}
\usepackage{epsfig}
\usepackage{epstopdf}
\usepackage{amsmath}
\usepackage{subfig}
\usepackage[dvipsnames]{xcolor}
\newcommand{\be}{\begin{equation}}
\newcommand{\ee}{\end{equation}}
\newcommand{\ba}{\begin{eqnarray}}
\newcommand{\ea}{\end{eqnarray}}
\newcommand{\nn}{\nonumber\\}

\begin{document}
\title{Collective excitations of hot QCD medium in a quasi-particle description }
\author{M. Yousuf Jamal}
\email{mohammad.yousuf@iitgn.ac.in}
\author{Sukanya Mitra}
\email{sukanyam@iitgn.ac.in}
\author{Vinod Chandra}
\email{vchandra@iitgn.ac.in}
\affiliation{Indian Institute of Technology Gandhinagar,  Gandhinagar-382355, Gujarat, India}

\begin{abstract}
Collective excitations of a hot QCD medium  are the main focus of the present article.
 The analysis is performed within semi-classical transport theory with isotropic and anisotropic momentum distribution functions 
 for the gluonic and quark-antiquark degrees of freedom that constitutes the  hot QCD plasma. The isotropic/equilibrium  momentum
distributions for gluons and quarks are based on a recent quasi-particle description of hot QCD equations of state. The anisotropic distributions 
are just the extensions of isotropic ones by stretching or squeezing them in one of the directions. The hot QCD medium effects in the model adopted
 here enter through the effective gluon and quark fugacities along with non-trivial dispersion relations leading to an effective QCD coupling constant. 
Interestingly,  with these distribution functions the tensorial structure of the  gluon polarization tensor in the medium turned out to be similar to the one 
for the non-interacting ultra-relativistic system of  quarks/antiquarks and gluons . The interactions mainly modify the Debye mass parameter and , in turn, the effective coupling in the 
medium. These modifications have been seen to modify the collective modes of the hot QCD plasma in a significant way.
\end{abstract}
\maketitle

\section{Introduction}
 Heavy ion collision experiments at RHIC, BNL and LHC, CERN have already 
observed liquid like hot QCD matter in experiments~\cite{expt_rhic, expt_lhc}, commonly termed as quark-gluon plasma (QGP).
As one of fundamental aspect of strong interaction force in nature,  {\it viz.}, the confinement, prohibits to observe this medium 
directly in the experiments, therefore, indirect probes have been suggested for the existence of the QGP in the experiments. 
Among them, two of the robust ones  are, {\it viz.}, collective flow and jet quenching (suppression of high $P_T$ hadrons)
 which have been observed both at RHIC and LHC, heavy-ion collision experiments establishing the near perfect picture and 
strongly coupled nature of the QGP. The QGP is suggested to possess a very tiny value for the shear viscosity to entropy density ratio
that realizes the above mentioned signatures with quite a appreciable degree of accuracy in the experiments. 
Theoretical considerations based on various approaches starting from kinetic theory to holographic theory also
 support a tiny value for the shear viscosity to entropy density ratio for the matter therein~\cite{Ryu, Denicol1}. 

Another important signature to sense the existence of the strongly coupled medium is the quarkonia suppression. The main cause of the 
quarkonia dissociation at high temperature and strong suppression in the yields of various quarkonia ($J/\Psi$, $\Upsilon$ {\it etc.}), 
is the color Debye  screening~\cite{Chu:1988wh} in the QGP medium and other related phenomena such as Landau damping ~\cite{Landau:1984}
and energy loss~\cite{Koike:1991mf}. Study of these  aspects of the QGP require exploration on the spectrum of the collective excitations of the QGP.  
These collective excitations carry quite crucial information about the equilibrated QGP and also provides a handle on the temporal evolution of the
 non (near)-equilibrated one. In the context QGP, both quark and gluonic collective modes need to be explored. 
The prime focus of this work is to understand the gluonic collective modes only, which are obtained 
in terms of the gluon-polarization tensor in the QGP medium. To investigate the fermionic modes (quark  and antiquark modes), quark self-energy 
has to be obtained in the medium which is beyond the scope of present work.

There have been several attempts to understand the collective
 behavior of hot QCD medium either within the semi-classical theory or HTL ~\cite{Mrowczynski:1993qm,Mrowczynski:1994xv,Mrowczynski:1996vh}
effective theory. In both the approaches, dispersion equations for the collective excitations 
are obtained. These equations depict  conditions for  the existence of solutions 
of the homogeneous equation of motion. In the case of hot QCD/ QGP, the equations of motion 
are the Yang-Mills equations of the chromodynamic field.  In classical approach,  the equations of motion depend on the chromodynamic
permeability, or chromodielectric tensor, that encodes the effect of the QGP medium. 
 In QFT, polarization tensor that enters into the gluon propagator, contains the dynamical information about the medium and
 the dispersion equations are obtained in terms of poles of the gluon propagator.
The kinetic theory approach  is done within a linear response analysis of classical (or semiclassical)
transport equations while the  QFT formulation consists of the standard  
perturbative method within the hard thermal loop approximation. 
These two approaches  turn out to be completely equivalent ~\cite{dm_rev1,dm_rev2,Mrowczynski:2000ed,Mrowczynski:2004kv} 
and the chromo-dielectric tensor can be expressed directly in terms of the polarization tensor and vice versa. 
The quantum effects  that are taken into account, enter through the quantum statistics of the plasma constituents only.
The plasmons which are the glunoic collective modes in the present case, in the case of isotropic plasmas, have 
been done long ago and can be seen in the text books~\cite{Landau:1984, Bellac:1996}. Notably, the effects of isotropic distributions enter only through the mass 
parameter (the Debye mass). The present analysis is the extension of these approaches for the interacting (realistic) equations of state for the QGP while 
also including the small but finite anisotropic effects  in the system.

 Like QED plasma, hot QCD plasma (in the abelian limit) do possess similar collective excitations ~\cite{Weibel:1959zz}.
The transverse and longitudinal collective plasma excitations are commonly known as plasmons. There has been various attempts to study them
both in the isotropic and anisotropic hot QCD medium ~\cite{Romatschke:2003ms,Romatschke:2004jh,Mrowczynski:2005ki,Arnold:2003rq}.
The  distributions functions for partons, used to study different aspects of QGP are
 discussed in ~\cite{Attems:2012js,Florkowski:2012as,Dumitru:2007hy,Martinez:2008di,Schenke:2006yp}. 
 In most of the studies, the momentum distributions that
 are crucial to obtain the whole plasma spectrum have been modeled by considering the QGP as free ultra-relativistic gas
 of quarks and gluons. In view of the strongly coupled nature of QGP, this need refinement by systematically including 
the hot QCD medium effects at the level of momentum distributions before one setup an effective transport equation.

Our analysis, methodologically, is an adaptation of the works of Carrington, Deja and Mr\'{o}wzy\'{n}ski~\cite{Carrington:2014bla} and Romatschke and 
Strickland~\cite{Romatschke:2003ms}. We extended their formalism for interacting QGP equations of state where the medium effects are encoded 
in the momentum distribution of  effective gluons and effective quarks employing a recently proposed quasi-particle model~\cite{chandra_quasi1,chandra_quasi2}.
Another important aspect of the ultra relativistic heavy-ion collisions is the momentum anisotropy. 
As the initial geometry (just after collision) is approximately almond shaped with only spatial anisotropy (no momentum anisotropy). Due to 
interactions, the system quickly thermalizes and because of different pressure gradients in different
directions, it expands anisotropically. The momentum anisotropy, that develops at the cost of spatial anisotropy during the 
expansion of the system, is present throughout the hydrodynamical expansion of the QGP.Therefore, inclusion of such anisotropic 
effects while modeling the QGP is inevitable. The modeling of anisotropic distributions for the gluons and quarks in our case is based on~\cite{Carrington:2014bla}, while genralizing the 
isotropic distributions obtained in~\cite{chandra_quasi1,chandra_quasi2}.

The collective modes are defined in terms of the gluon propagator which, in turn,  depend on the gluon polarization in the medium. In the case of isotropic QGP ,
we can define modes as longitudinal ones and transverse ones. Generally, we define a longitudinal mode (chomo-electric mode) when the chromoelectric field 
is parallel to wave vector $k$. On the other hand a  mode is called transverse  (chromo-magnetic mode),
when the chromoelectric field is transverse to the wave vector. The Maxwell's equations (classical Yang-Mills equations in the case of the QGP) show that 
longitudinal modes are associated with chromo-electric charge oscillations and the transverse modes
are associated with chromo-electric current oscillations. Both of them can be   stable or unstable  and can be determined as a solution
of the dispersion relations which can be obtained from the poles of the gluon propagator. 
Within linear response theory, we compute the $\Pi^{\mu \nu}$
(gluon-polarization tensor) from semi-classical transport theory consideration.
Thereby, obtained an expression for the chromo-electric pemittivity tensor invoking 
classical Yang-Mills equations in the abelian limit. On the other hand, the anisotropic QGP develop the tensorial response and different modes in the 
different directions. Interestingly, presence of anisotropy leads to both real and imaginary modes in the QGP medium.
 
The paper is organized as follows. In section II, the basic formalism of gluon polarization tensor
has been presented along with the modeling of hot QCD medium for isotropic as well as small anisotropic hot QCD medium 
through different sub-sections. Section III deals with the brief mathematical formalism for different collective modes along 
with the results and discussions. Summary and conclusions of the present work is offered in Section IV.

\section{Gluon polarization tensor in  QCD plasma}
To understand  the properties of any plasma medium, 
investigations on the structure of its polarization tensor, 
 is an essential exercise. In the context of the QGP/hot QCD medium, 
gluon polarization tensor, $\Pi^{\mu\nu}$ is the main focus of our interest.
The $\Pi^{\mu\nu}$ contains all the information about the medium as it 
 describes the interaction term in the effective Lagrangian.
 The gluon polarization
 tensor in our case is obtained employing semi-classical kinetic theory approach to the QGP. This 
 requires equilibrium/isotropic  modeling of hot QCD medium first. If one is dealing with the
 anisotropic case, the near equilibrium distributions  also need to be known for the gluons and quarks
  to start up the analysis.
  
 As the present analysis deals with abelian responses  (the singlet part
 of the gluon self-energy/polarization), therefore, the  framework that is already available for the 
 QED plasma make sense here.  Remember that while dealing with hot QCD plasma the   extra 
 degree of freedom {\it i.e.}, the color degree of freedom need to  be appropriately incorporated in the analysis.
 In QCD plasma, we  start with an arbitrary particle distribution function, 
denoting with $f_{i}({\bf p},x)$ where index $i$ labeling the $i$- th particle. 
The function depends on the space-time $x(t,{\bf x})$ and the three-momentum ${\bf p}$.
The four-momentum $p$ obeys the mass-shell constraint $p^{2} = m^{2}$ , where $m$ is the
 mass of the particle ( $p \equiv(E_{\bf p} , {\bf p})$ 
with $E_{\bf p} \equiv m ^{2} + {\bf p}{2}$) .

The space time evolution of the distribution function in the medium  is understood from 
the Boltzmann-Vlasov ~\cite{dm_rev2, Elze:1989un} transport equation below.
\begin{equation}
 p^{\mu}\partial_{\mu}f_{i} - q_{i} F_{\mu\nu}p^{\mu} \frac{\partial f_{i}}{\partial p_{\nu}} = C[f_{i}],
 \label{eq:Vlasov}
\end{equation}
where  $q_{i}$ is the charge of the plasma species
$i$, and $C[f_{i}]$ denotes the collision term.  The near equilibrium considerations in our analysis 
allows us to neglect the effects from collisions and so  $C[f_{i}] = 0$. The second rank tensor, 
$F_{\mu\nu}$ is the electromagnetic strength tensor which either represents an external field 
applied to the system, or/and is generated self-consistently by the four-currents present in the  plasma, as follows,
\begin{equation}
 \partial_{\nu}F^{\mu\nu} = 4\pi j^{\mu}
\end{equation}
where
\begin{equation}
 j^{\mu}(x) = \varSigma_{i}q_{i}\int d \varGamma p^{\mu} f_{i}({\bf p},x)
 \end{equation}
 and
\begin{equation}
  d\varGamma \equiv \frac{d^{3}p}{(2\pi)^{3}E} .
 \end{equation}
The transport equation, Eq.(\ref{eq:Vlasov}), can be solved in the linear response approximation.
The equation is linearized around the stationary and homogeneous state described
by the distribution $f_{i}({\bf p},x)$. The state is also assumed to be neutral with the absence of 
any current. The distribution function is then decomposed as:
 \begin{equation}
   f_{i}(x) = f^{0}_{i}({\bf p}) + \delta f_{i}({\bf p}, x)    
 \end{equation}
 Where 
 \begin{equation}
   f^{0}_{i}({\bf p}) \gg \delta f_{i}({\bf p}, x)    
 \end{equation}
 So the induced polarization current will be
 \begin{equation}
 j^{\mu}_{ind}(x) = \varSigma_{i}q_{i}\int d\varGamma p^{\mu}\delta f_{i}({\bf p},x) 
 \end{equation}

Next, in order to study the collective behaviour of hot QCD plasma one needs to choose a  
particular energy scale. Since the collective motion in the hot medium first appears at the soft momentum scale, 
$p \sim g T \ll T$,  so the main focus will be around that only.
The magnitude of the field fluctuations is of the order of $A \sim \sqrt{g} T$ and derivatives are $\partial_x \sim g T$ at the said scale. 
Now, following the analysis of Romatschke and Strikland~\cite{Romatschke:2003ms}, a covariant gradient expansion of the quark and gluon Wigner function in the mean-field approximation
can be performed at the leading order in the coupling constant.  The color current, $j^{\mu,a}$, induced by a soft gauge field, 
$A^{\mu}$, with four-momentum $k=(\omega,{\bf k})$ can be obtained as follows,
\begin{equation}
j^{\mu,a}_{\rm ind}(x) =  g \int  d\varGamma p^{\mu} \delta f^{a}({\bf p}, x)
\end{equation}
here $\delta f^{a}({\bf p},x)$  contains the fluctuating part given as

\begin{equation}
\begin{split}
 \delta f^{a}({\bf p},x) = 2 N_{c} \delta f_{g}^{a}({\bf p},x) + N_{f} (\delta f_{q}^{a}({\bf p},x)  \\
 - \delta f_{\bar{q}}^{a}({\bf p},x)).
\end{split}
\end{equation}
 Where $\delta f_{g}^{a}({\bf p},x)$, $\delta f_{q}^{a}({\bf p},x)$ and $f_{\bar{q}}^{a}({\bf p},x)$ 
are the fluctuating parts of the gluon, the quark and anti-quark densities, respectively.  Note that, 
$\delta f^{a}_{g}({\bf p},x)$ transforms as a vector in the adjoint representation ($\delta f_{g}({\bf p},x) \equiv \delta f^{a}_{g}({\bf p},x)T^{a}$)
 where as $\delta f^{a}_{q/\bar{q}}({\bf p},x)$ transforms as a vector in the fundamental representation ($\delta f_{q/\bar{q}}({\bf p},x) \equiv \delta f_{q/\bar{q}}^{a}({\bf p},x)t^{a}$).

These quark and gluon density matrices satisfy the following transport equations:
\begin{equation}
\bigg[ u \cdot D_{x}, \delta f_{q/\bar{q}}({\bf p},x) \bigg] =   \mp g u_{\mu} F^{\mu\nu}({\bf p},x) \partial_{\nu} f_{q/\bar{q}}({\bf p}) 
\end{equation}
\begin{equation}
\bigg[ u \cdot D_{x}, \delta f_{g}({\bf p},x) \bigg] = - g u_{\mu} F^{\mu \nu}({\bf p},x) \partial_{\nu} f_{g}({\bf p})  
\end{equation}
where $D_x = \partial_x + i g A(x)$ is the covariant derivative. After solving the transport equations for the fluctuations 
$\delta f_{g}$ and  $\delta f_{q/\bar{q}}$,  we get the induced current as
\ba
 j^{\mu}_{\rm ind}(x) &=& g^2 \int \frac{d^3 p}{(2\pi)^3} u^{\mu} u^{\alpha}
	\partial^{\beta}_{(p)} f({\bf p})\nn&& \int d\tau U(x,x-u\tau) F_{\alpha \beta}(x-u\tau)\nn&& U(x-u\tau,x),
 \ea
where $U(x,y)$ is a gauge parallel transporter defined by the path-ordered integral
\begin{equation}
U(x,y) = \mathcal{P} \, {\rm exp}\left[ - i g \int_x^y d z_\mu A^\mu(z) \right] \; ,
\end{equation}
and
\begin{equation}
F_{\alpha \beta} = \partial_\alpha A_\beta - \partial_\beta A_\alpha - i g [A_\mu,A_\nu]
\end{equation}
is the gluon field strength tensor and

\begin{equation}
\label{fq}
f({\bf p}) = 2 N_c f_{g}({\bf p})+ N_{f} (f_{q}({\bf p})+f_{\bar{q}}({\bf p})).
\end{equation}
Here, $f_{q,g} ({\bf p})$ denote the isotropic/equilibrium quark/antiquark (at zero baryon density) and gluon distribution functions. To obtain them for interacting hot QCD medium, we need to 
adopt effective description of hot QCD equations of state. This point will be discussed  later.

After neglecting terms that are subleading order in $g$ (implying $U\rightarrow1$ and $F_{\alpha \beta} 
\rightarrow \partial_\alpha A_\beta - \partial_\beta A_\alpha$) 
one can find $\delta f_{i}({\bf p}, k)$  {\it i.e.}, the Fourier transform of $\delta f_{i}(p, x)$ and can obtain the Fourier transform of the 
induced current as
\begin{equation}
\begin{split}
 j'^{\mu}_{\rm ind}(k) = g^2 \int \frac{d^3 p}{(2\pi)^3} u^{\mu}
\partial^{\beta}_{(p)} f({\bf p}) \bigg[ g_{\alpha \beta} -
\\ \frac{u_{\alpha} k_{\beta}}{k\cdot u + i \epsilon}\bigg] A'^{\alpha}(k),
\end{split}
\label{eq:induced current}
\end{equation}
where $\epsilon$ is a small parameter just to take care of unwanted infinities and will be sent to zero in the end.
In the linear approximation the equation of motion for the gauge field can be obtained in Fourier space as,
\begin{equation}
     j'^{\mu}_{ind}(k) = \Pi^{\mu\nu}(k)A'_{\nu}(k). 
     \label{eq:linear induced current}
 \end{equation}
 Here, $\Pi^{\mu\nu}(k) = \Pi^{\nu\mu}(k)$ and follows the Ward's identity, $k_{\mu}\Pi^{\mu\nu}(k) = 0$, i.e., the self-energy tensor is symmetric and transverse in nature.
From Eq.(\ref{eq:induced current}) and Eq.(\ref{eq:linear induced current}) we can obtain $\Pi^{\mu\nu}(k)$ as  
 \begin{equation}
   \Pi^{\mu\nu}(k) = g^{2}\int \frac{d^{3}p}{(2\pi)^{3}} u^{\mu}\frac{\partial f(p)}{\partial p^{\beta}}\bigg[g^{\nu\beta} - \frac{u^{\nu}k^{\beta}}{u.k + i\epsilon}\bigg].
   \label{iso_pi}
\end{equation}    
The quantity,  $u^{\mu} \equiv (1 , {\bf k}/|{\bf k}|)$ is a light-like vector describing the propagation of plasma particle in space-time and
 \begin{equation}
g_{\mu\nu} = diag(1,-1,-1,-1)
 \end{equation}
is Minkowski metric tensor. 

Putting $j'^{\mu}_{ind}(k)$ and $F'^{\mu\nu}(k)$ into following Maxwell's equation
\begin{equation}
-ik_{\mu} F'^{\mu\nu}(k) = j'^{\nu}_{\rm ind}(k) + j'^{\nu}_{\rm ext}(k)  ,
\end{equation}
we get
\begin{equation}
\bigg[k^{2} g^{\mu\nu} - k^{\mu} k^{\nu} + \Pi^{\mu\nu}(k)\bigg]A'_{\mu}(k) = - j'^{\nu}_{\rm ext}(k), 
\label{eq:ext current}
\end{equation}
where $j'^\nu_{\rm ext}(k)$ is an external current in the  Fourier space.

Using the temporal axial gauge defined by $A_0=0$( with $ A^{j} = \frac{E^{j}}{i\omega}$), we can
write Eq.(\ref{eq:ext current}) in terms of a physical electric field as  
\begin{equation} 
\begin{split}
\bigg[(k^2-\omega^2)\delta^{ij} - k^{i} k^{j} + \Pi^{ij}(k)\bigg] E^{j}(k)\\ = i \omega j'^{i}_{\rm ext}(k) 
\end{split}
\end{equation}
Re-writing the above equation in the form of propagator,
\begin{equation} 
  [\Delta^{ij}(k)]^{-1} E^j(k) = i \omega j_{\rm ext}^i(k).
  \label{eq:propagator}
  \end{equation}
where 
\begin{equation} 
\Delta^{ij}(k) = \frac{1}{(k^2-\omega^2)\delta^{ij} - k^{i} k^{j} + \Pi^{ij}(k)},
\label{eq:propagator}
  \end{equation}
we can determine the response of the system to the external source,
\begin{equation}
E^j(k) = i \omega \, \Delta^{ij}(k) j_{\rm ext}^i(k) \; .
\end{equation}
The poles in the propagator $\Delta^{ij}(k)$ in Eq.(\ref{eq:propagator}) will lead to the dispersion relations for the collective modes.
Below, we shall discuss about the gluon 
polarization tensor for isotropic case as well as in small anisotropy($\xi$) limit.

\subsection{Isotropic hot QCD medium}
In the case of isotropic hot QCD medium, we work with 
 an effective model that describes hot QCD medium effects in terms of non-intercating/weakly interacting quasi-particle quarks and gluons.
 Here, hot QCD medium effects encoded in the equation of state have been described in terms of effective quasi-particle degrees of freedom 
within various different approaches. There  are several quasi-particle descriptions {\it viz.},
 effective mass  models~\cite{effmass1, effmass2}, effective mass models with Polyakov loop~\cite{polya}, 
NJL and PNJL based effective models~\cite{pnjl}, and effective fugacity quasi-particle description of hot QCD (EQPM)~\cite{chandra_quasi1, chandra_quasi2}.
 The present analysis considers  the  EQPM for the investigations on the properties of hot and dense medium  in RHIC.  

These quasi-particle models have shown their utility while  studying transport properties of the QGP~\cite{Bluhm,chandra_eta, chandra_etazeta}.
In  Ref. ~\cite{Bluhm},  shear and bulk viscosities ($\eta$ and $\zeta$)   for pure glue plasma have been estimated  employing the effective mass  model. 
In Refs. ~\cite{chandra_eta, chandra_etazeta},  these are estimated both in gluonic as well as matter sector.  A general quasi-particle theory of
 $\eta$ and $\zeta$ in the hadronic sector has been presented in~\cite{PJI,Mkap}. The authors also estimated them for hadronic sector. Further, thermal conductivity
has also been studied, in addition to the viscosities~\cite{Mkap}, again within the effective mass model.
In  Ref.~\cite{Greco}, the ratio of electrical conductivity to shear viscosity has been explored 
within the framework of the same quasiparticle approach as well. In a very recent work,  Mitra and  Chandra~\cite{Mitra:2016zdw}
estimated the electrical conductivity and charge diffusion coefficients employing EQPM. The EQPM has also served as 
basis to incorporate hot QCD medium effects while investigating heavy-quark transport in isotropic~\cite{Das:2012ck} and anisotropic 
~\cite{Chandra:2015gma} along with quarkonia in hot QCD medium~\cite{Chandra:2010xg, Agotiya:2016bqr}  and thermal particle
 production~\cite{Chandra:2015rdz,Chandra:2016dwy}. However,  the above model calculations were not  able to exactly yield  the shear and bulk
viscosities phenomenologically extracted from the hydrodynamic simulations of the QGP \cite{Ryu, Denicol1}, consistently agreeing with different
experimental observables measured such as  the multiplicity, transverse momentum spectra and the integrated flow harmonics of charged hadrons.
Nevertheless, these quasiparticle approaches  play crucial role the isotropic modeling of the QGP medium.
 
The EQPM employed here, models the 
hot QCD in terms of effective quasi-gluons and quasi-quarks/antiquarks with respective  effective fugacities. The idea has been to  map the 
hot QCD medium effects present in the equations of state (EOSs) either computed within improved perturbative QCD or lattice QCD simulations, 
in terms of effective equilibrium distribution functions for the quasi-particles.  Three  different EOSs and their respective EQPM descriptions denoted by EOS1, EOS2 and LEOS have been considered here. Noteworthily, the EOS1 
is fully perturbative and computed up to  $O(g^5)$ by Arnold and Zhai~\cite{zhai} and Zhai and Kastening~\cite{kastening} and 
EOS2 which is up to  $O(g^6\ln(1/g)+\delta)$ is  computed by Kajantie {\it et al.}~\cite{kaj}  while
 incorporating contributions from non-perturbative scales such as  $g T$ and $g^2 T$. 
The LEOS is the recent (2+1)-flavor lattice QCD EOS~\cite{cheng} at physical quark masse. 
Note that there are more recent lattice results with the improved 
actions and refined lattices~\cite{leos1_lat}, for which we need to re-look the model 
with specific set of lattice data specially to define the effective gluonic degrees of freedom. This is beyond the scope of present analysis.
Therefore, we will stick to  LEOS and the model for it that is described in Ref.~\cite{chandra_quasi2}. However, in view of the fact that the quasi-particle model is 
more reliable beyond $1.5 T_c$, the changes that are expected are perhaps not so major.

Next,  the form of the quasi-parton equilibrium distribution functions (describing either of the EOSs)
 $ f_{eq}\equiv \lbrace f_{g}, f_{q} \rbrace$  (describing the strong interaction effects in terms of effective fugacities $z_{g,q}$) can be written as,
\be
\label{eq1}
f_{g/q}= \frac{z_{g/q}\exp[-\beta E_p]}{\bigg(1\mp z_{g/q}\exp[-\beta E_p]\bigg)}
\ee
where $E_p=|\vec{p}|$ for the gluons and $\sqrt{|\vec{p}|^2+m_q^2}$ for the quark degrees of freedom ($m_q$ denotes the mass of the quarks).
Since the model is valid in the deconfined phase of QCD (beyond $T_c$), therefore, the mass of the light quarks can be neglected as compared to 
the temperature. Notably,  these effective fugacities ($z_{g/q}$)  lead to 
non-trivial dispersion relation both in the gluonic and quark sectors:
\ba
\label{eq2}
\omega_{g/q}=E_p+T^2\partial_T ln(z_{g/q}).
\label{epp}
\ea
The second term in the right-hand side of Eq.(\ref{eq2}), encodes the effects from collective excitations of the quasi-partons.
Note that   $z_g, z_q$ are not related with any conserved number current in 
the hot QCD medium and they are introduced to  encode interaction effects.  Note that $f_{\bar{q}}$ in Eq. (\ref{fq}) has exactly  same expression  $f_{q}$  (zero baryon density ).
Note that, extension of the EQPM, while incoporating the non-extensive quantum statistics, has recently been done by 
Rozynek and Wilk~\cite{Rozynek:2016ykp}. We intend to utilize these non-extensive quasi-particle
distribution in the context of collective plasma excitations of hot QCD medium in near future.

\subsubsection{Debye mass and effective QCD coupling}
Next, we employ  the EQPM to obtain the Debye mass and effective coupling in hot QCD medium.
The mathematical expression for the Debye mass, derived in semi-classical transport theory~\cite{dmass1, dm_rev1, dm_rev2} given below in terms of 
equilibrium gluonic and quark/antiquark distribution function can be employed here, 
 \ba
 \label{dm}
 m_D^2&=& -4 \pi \alpha_{s}(T) \bigg(2 N_c \int \frac{d^3 p}{(2 \pi)^3} \partial_p f_g (\vec{p})\nn
 &+& 2 N_f  \int \frac{d^3 p}{(2 \pi)^3} \partial_p f_q (\vec{p})\bigg),
 \ea
 where, $\alpha_{s}(T)$ is the QCD running coupling constant at finite temperature~\cite{qcd_coupling}. 
 Employing EQPM, we obtain the following expression,
 \ \ba
 \label{dm1}
 {m_D^2}^{EOS(i)}&=&4 \pi \alpha_{s}(T) T^2  \bigg( \frac{2 N_c}{\pi^2} PolyLog[2,z_g]\nn&-&\frac{2 N_f}{\pi^2} PolyLog[2,-z_q]\bigg). 
 \ea
 Where $i =1$ (EOS1), $i=2$ (EOS2) and $i=2+1$ (LEOS)  ($z_{g,q}$) computed from respective EOSs)
 In the  limit ( $z_{g,q}=1$)  the $m_D$ reduces to the leading order (LO)
 expression (ideal  EOS: non-interacting of ultra relativistic quarks and gluons),
 \be
 {m_D^2}^{LO}= \alpha_{s}(T) \ T^2 (\frac{N_c}{3}+\frac{N_f}{6}).
\ee
Eq.(\ref{dm1}) can be rewritten as, 
\ba
\frac{{m_D^2}^{EOS(i)}}{{m_D^2}^{LO}}&=& \frac{\frac{2 N_c}{\pi^2} PolyLog[2,z_g]-\frac{2 N_f}{\pi^2} PolyLog[2,-z_q]}{\frac{N_c}{3}+\frac{N_f}{6}}.\nn
\label{eq:md_comp}
\ea
Now, defining the effective QCD coupling, 
\ba
\alpha_{eff} \equiv \alpha_{s}(T) g(z_g,z_q) 
\ea
we obtain:
\ba
{m_D^2}^{EOS(i)}& =& 4\pi \alpha_{eff} (T)\  T^2 (N_c/3+N_f/6).  
\ea
 The function $g(z_g,z_q)$,reads,
\ba
g(z_g,z_q)&=& \frac{\frac{2 N_c}{\pi^2} PolyLog[2,z_g]-\frac{2 N_f}{\pi^2} PolyLog[2,-z_q]}{\frac{N_c}{3}+\frac{N_f}{6}}
\ea
It is important to note that with EQPM employed here, due to the isotropic nature of the 
effective  momentum distribution functions of quarks and gluons, the mathematical structure of the analysis remains intact to the 
formalism of~\cite{Romatschke:2003ms}. The temperature dependence of the Debye mass parameter that captures the effects from effective coupling is 
different in our case and in the limiting case (Stefan-Boltzmann limit, $z_{g,q} \rightarrow 1$), it yields the same expression as employed in~\cite{Romatschke:2003ms}.

Now, in order to solve that integral shown in Eq.(\ref{iso_pi}) in the isotropic case, we construct a co-variant form of $\Pi^{\mu\nu}$ analytically by
making the possible combination of available symmetric tensors. The anlysis is similar as done in \cite{Liu:2011if} which are $[g_{\mu\nu}, k_{\mu}k_{\nu}, u_{\mu}u_{\nu},k_{\nu}u_{\mu} + k_{\mu}u_{\nu}]$. 
One of the possible combination to construct $\Pi^{\mu\nu}(k)$ in isotropic case can be obtained by using,
\begin{eqnarray}
P_{\mu\nu} & = & g_{\mu\nu}-u_{\mu}u_{\nu}+\frac{1}{{\bf k^{2}}}\left(k_{\mu}-\omega u_{\mu}\right)\left(k_{\nu}-\omega u_{\nu}\right),\nonumber \\
Q_{\mu\nu} & = & \frac{-1}{{\bf k^{2}}k^{2}}\left(\omega k_{\mu}-k^{2}u_{\mu}\right)\left(\omega k_{\nu}-k^{2}u_{\nu}\right).\label{eq:projectors}\end{eqnarray}
as
\begin{equation}
\Pi_{\mu\nu}(k)=\Pi_{T}(k)P_{\mu\nu}+\Pi_{L}(k)Q_{\mu\nu},\label{eq:PQ}\end{equation}
where $\Pi_{T}(k)$ is the transverse and $\Pi_{L}(k)$ is the  longitudinal part of the self-energy.
Using the following contractions,
  \begin{eqnarray}
  (g_{\mu\nu}-u_{\mu}u_{\nu})\Pi^{\mu\nu}(k) = 2\Pi_{T}(k) + (1+\frac{{\bf k^{2}}}{k^{2}})\Pi_{L}(k)
  \end{eqnarray}
and\begin{eqnarray}
u_{\mu}\Pi^{\mu\nu}u_{\nu} = -\Pi_{L}(k) \frac{{\bf k^{2}}}{k^{2}}
\end{eqnarray}

The form of $\Pi_{T}(k)$ and $\Pi_{L}(k)$  can be obtained as follows:
\begin{equation}
\Pi_T(k) = {m^{2}_D\over2} {\omega^2 \over k^2} \bigg[1-{\omega^2 - k^2 \over 2 \omega k}\log{\omega+k\over\omega-k}\bigg] 
\label{eq:Pi_T}
\end{equation}
\begin{equation}
\Pi_L(k) = m_D^2 \bigg( 1- \frac{\omega^2}{k^2} \bigg)\left[1- {\omega\over 2k} \log{\omega+k\over\omega-k}\right]  
\label{eq:Pi_L}
\end{equation}
For real valued $\omega$, $\Pi_T(k)$ and  $\Pi_L(k)$ are real for all $\omega > k$ and complex for $\omega < k$. Since for the
real valued $\omega$ the quantity inside the $\log$ is give as
\begin{equation}
 \log\bigg(\frac{\omega + k}{\omega - k}\bigg) = \log\bigg(\bigg|\frac{k + \omega}{k - \omega}\bigg|\bigg) - i\pi\varTheta\big(k - \omega\big)
 \label{eq:log}
\end{equation}
where $\varTheta$ is the heavyside theta function.

\subsection{Extension to anisotropic hot QCD medium}
To describe the anistropic hot QCD medium, we follow the approach of ~\cite{Romatschke:2003ms} and for the 
normalization of the momentum distribution function, we follow~\cite{Carrington:2014bla}.
In these approaches,  the anisotropic momentum distribution functions for the gluons and quark-antiquarks 
are done by rescaling (stretching and squeezing) of only one direction in momentum space function as:
\ba
	  f_{\xi}({\bf\tilde{p}}) = C_{\xi}f(\sqrt{{\bf p}^{2} + \xi({\bf p}.{\bf \hat{n}})^{2}})
	  \label{aniso_distr}.
 \ea
This introduces  one more degree of freedom,  {\it viz. }, the direction of anisotropy, ${\bf \hat{n}}$ with ${\bf \hat{n}}^{2} = 1$.
The anisotropy parameter   $\xi$  can be adjusted to reflect either squeezing or stretching   ({\it i . e.}  $\xi > 0$ correspond to contraction 
 of the distribution in the ${\bf \hat{n}}$ direction  or   $-1<\xi<0$ correspond to stretching of the distribution
 in the $\hat{n}$- direction) and $C_{\xi}$ is the normalization constant. Choosing, $\xi = 0$ will take  back to the case of isotropic medium. 

Next,  a change of 
 variables enable us to to integrate out the 
$(|\tilde p|\equiv \sqrt{{\bf p}^{2} + \xi({\bf p}.{\bf \hat{n}})^{2}} )$  in   $\Pi^{ij}$ in  Eq. (18) as,
\begin{equation}
\Pi^{ij}(k) = -g^2\int\frac{d^3p}{(2\pi)^3} u^{i}\partial^{l}f_{\xi}({\bf \tilde{p}})\left( \delta^{j l} + \frac{u^{j} k^{l}}{k\cdot u+i0^{+}}\right) 
  \label{eqpan}
\end{equation}
Again, a tensor decomposition of $\Pi^{\mu\nu}$ for an anisotropic system is possible to
develop in which there is only one preferred direction {\textit i.e.,} the direction of 
anisotropy, ${\bf \hat{n}}$. Since the self-energy is symmetric and transverse, all
components of $\Pi^{\mu\nu}$ are not independent. Therefore,  we can restrict our 
considerations to the spatial part of $\Pi^{\mu\nu}$, denoted as $\Pi^{ij}$ and
following the analysis done in ~\cite{Romatschke:2003ms,Kobes:1990dc,Carrington:2014bla},
we have:
\begin{equation}
\Pi^{ij} = \alpha A^{ij} + \beta B^{ij} + \gamma C^{ij} + \delta D^{ij},
 \end{equation}
where
\ba
 A^{ij} &=& \delta^{ij} - k^{i}k^{j}/k^{2}, \  \
 B^{ij} = k^{i}k^{j}/k^{2} \nn
C^{ij} &=& \tilde{n}^{i}\tilde{n}^{j}/\tilde{n}^{2},\  \
 D^{ij} = k^{i}\tilde{n}^{j}+ \tilde{n}^{i}k^{j}. 
 \ea
Using this tensor decomposition, $\Pi^{ij}$ is decomposed into four structure functions 
These are determined by considering the following contractions,
 \ba
k^{i}\Pi^{ij}k^{j} &=& k^{2}\beta,  \  \
 \tilde{n}^{i}\Pi^{ij}k^{j} = \tilde{n}^{2}k^{2}\delta \nn 
  \tilde{n}^{i}\Pi^{ij}\tilde{n}^{j} &=&\tilde{n}^{2}(\alpha + \gamma), \  \
  Tr\Pi^{ij} = 2\alpha +\beta + \gamma. 
  \label{pi_ij_aniso}
  \ea
and in terms of related  scalar quantities (structure functions), $\alpha$, $\beta$, $\gamma$ and $\delta$. Here,  $\tilde{n}^{i} = A^{ij}{n}^{j}$ which obeys $\tilde{n}\cdot k = 0$.

Note that,  all of the  four structure functions depend on $m_D$,  $\omega$, $k$, $\xi$, and
${\bf \hat k}\cdot{\bf \hat n}=\cos\theta_n$. In the limit of vanishing strength of the anisotropy,  $\xi \rightarrow 0$, 
 the structure functions $\alpha$ and $\beta$  reduce to the isotropic hard-thermal-loop self-energies and $\gamma$ and $\delta$ just
vanish.
\ba
\alpha_{iso}(\omega,{\bf k}) &=& {m^{2}_D\over2} {\omega^2 \over k^2} \bigg[1-{\omega^2 - k^2 \over 2 \omega k}\log{\omega+k\over\omega-k}\bigg]  = \Pi_T(k) \nn
\beta_{iso}(\omega,{\bf k}) &=&  m_D^2 \frac{\omega^2}{k^2} \left[{\omega\over 2k} \log{\omega+k\over\omega-k} - 1\right]  \approx \Pi_L(k) \nn
\gamma_{iso}(\omega,{\bf k}) &=& 0,\   \   \ \delta_{iso}(\omega,{\bf k}) = 0
\label{coe_iso}
\ea

In the case of massless partons, the whole spectrum of collective excitations depends on a single mass 
parameter which is nothing but the Debye mass given in Eq.(\ref{dm}). In order to compare the collective
modes at different anisotropy , we have to make this parameter independent of $\xi$ but of course the effects
of different EOSs will be there in $m_D$.  While keeping the Debye mass intact with respect to $\xi$ in both the isotropic as
well as in the anisotropic case, we found the normalization constant($C_{\xi}$) to be~\cite{Carrington:2014bla},
\begin{equation}
 C_{\xi} = \frac{\sqrt{|\xi|}}{arctan\sqrt{|\xi|}}.
 \label{aniso_const}
\end{equation}
which is in small $\xi$ limit given as,
\begin{equation}
 C_{\xi} = 1 +\frac{\xi}{3} +O({\xi^2}).
 \label{small_aniso_const}
\end{equation}
the distribution shown in Eq.(\ref{aniso_distr}) is weekly stretched and squeezed for  $\xi < 0$ and $\xi > 0$ respectively.

\subsection{Structure functions in small anisotropy, $\xi$ limit}
 \begin{figure*}[]
  \centering
\subfloat{\includegraphics[height=5cm,width=6cm]{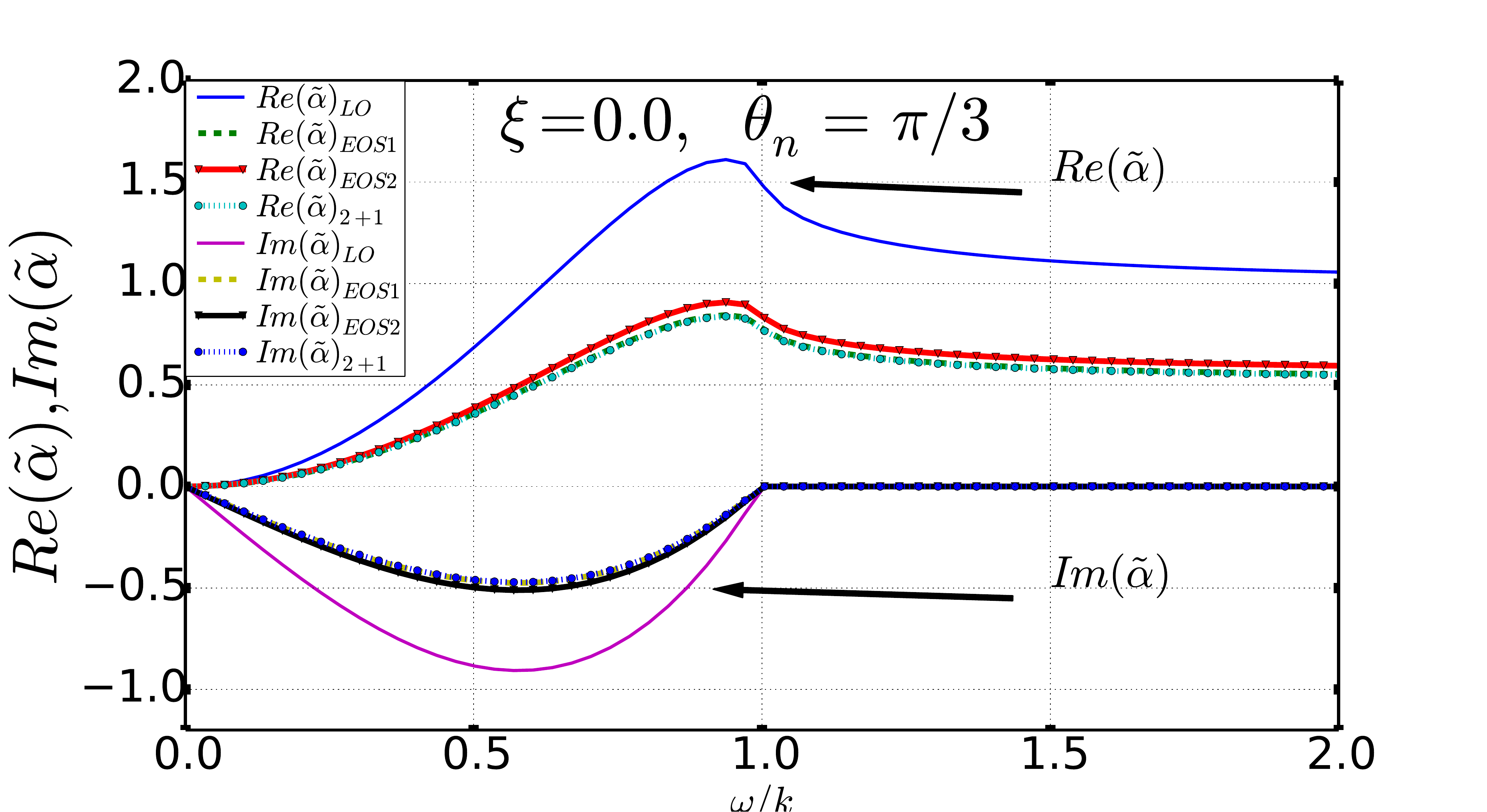}}
\subfloat{\includegraphics[height=5cm,width=6cm]{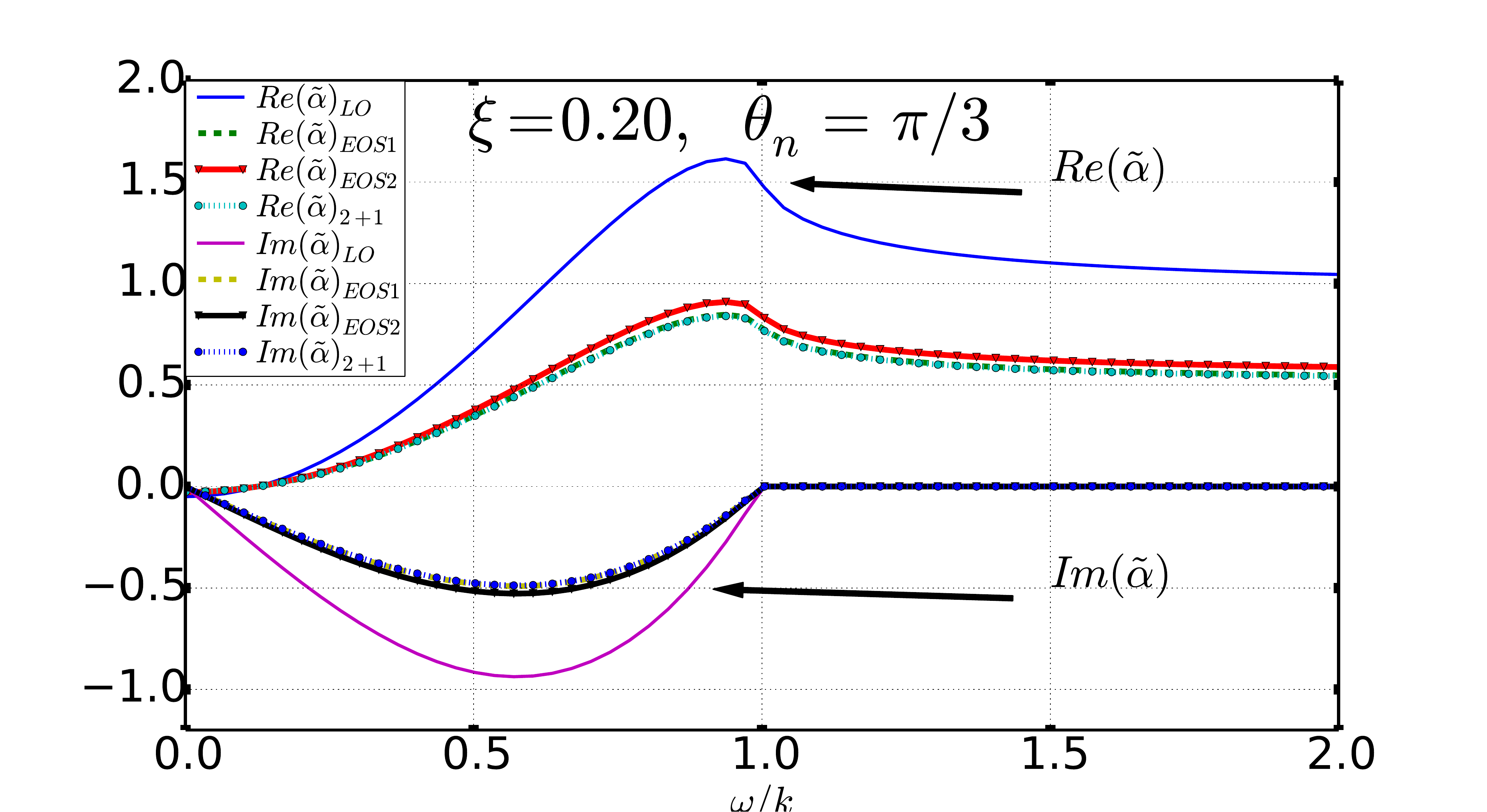}}
\subfloat{\includegraphics[height=5cm,width=6cm]{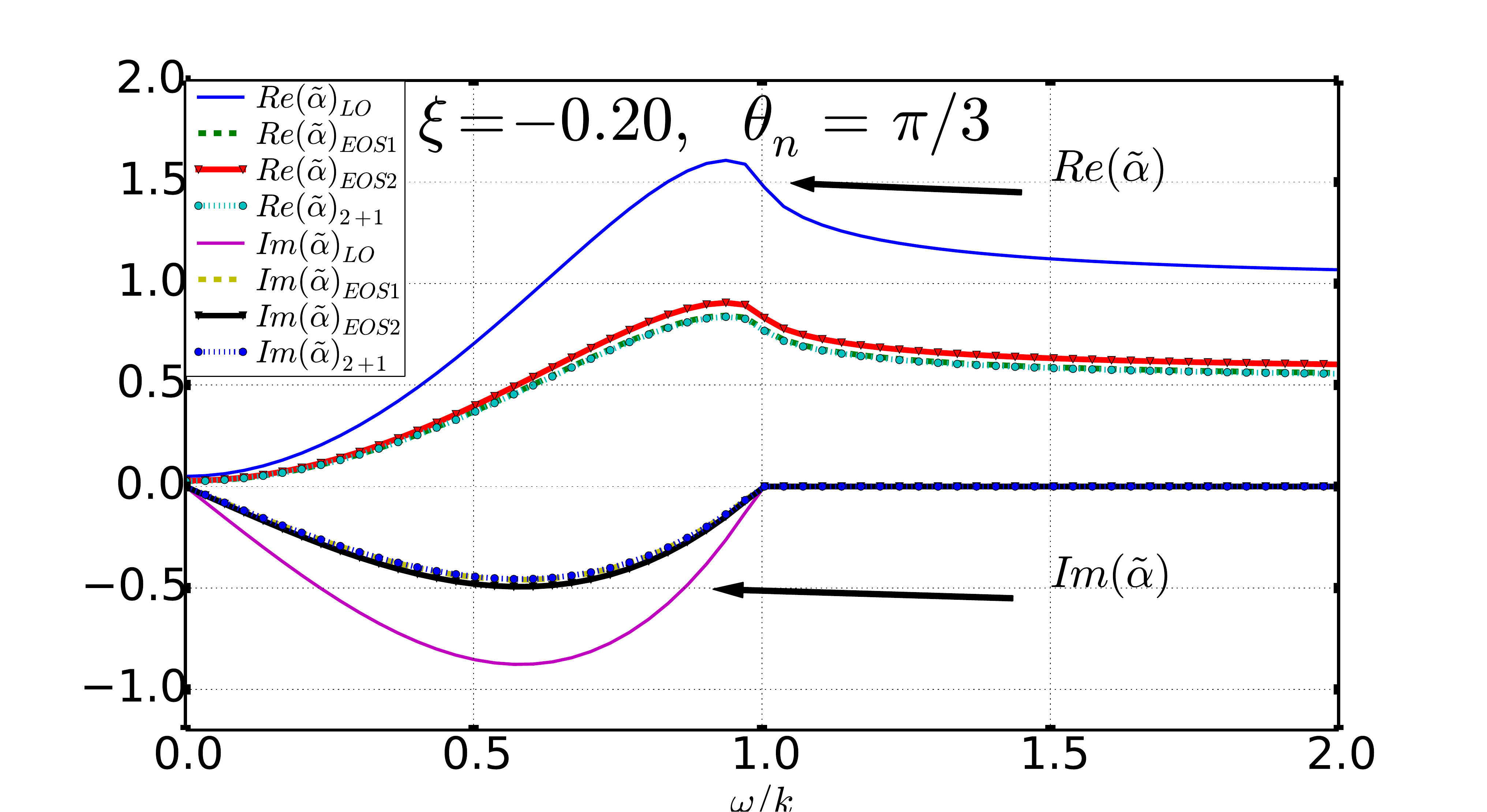}}
\caption{(Color online) $Re(\tilde{\alpha})$ and $Im(\tilde{\alpha})$ with $\omega/k$ at $T_{c} = 0.17GeV$ and $T = 0.25$ for various EOSs.}
\label{fig:alpha}
\end{figure*}
\begin{figure*}
  \centering
\subfloat{\includegraphics[height=5cm,width=6cm]{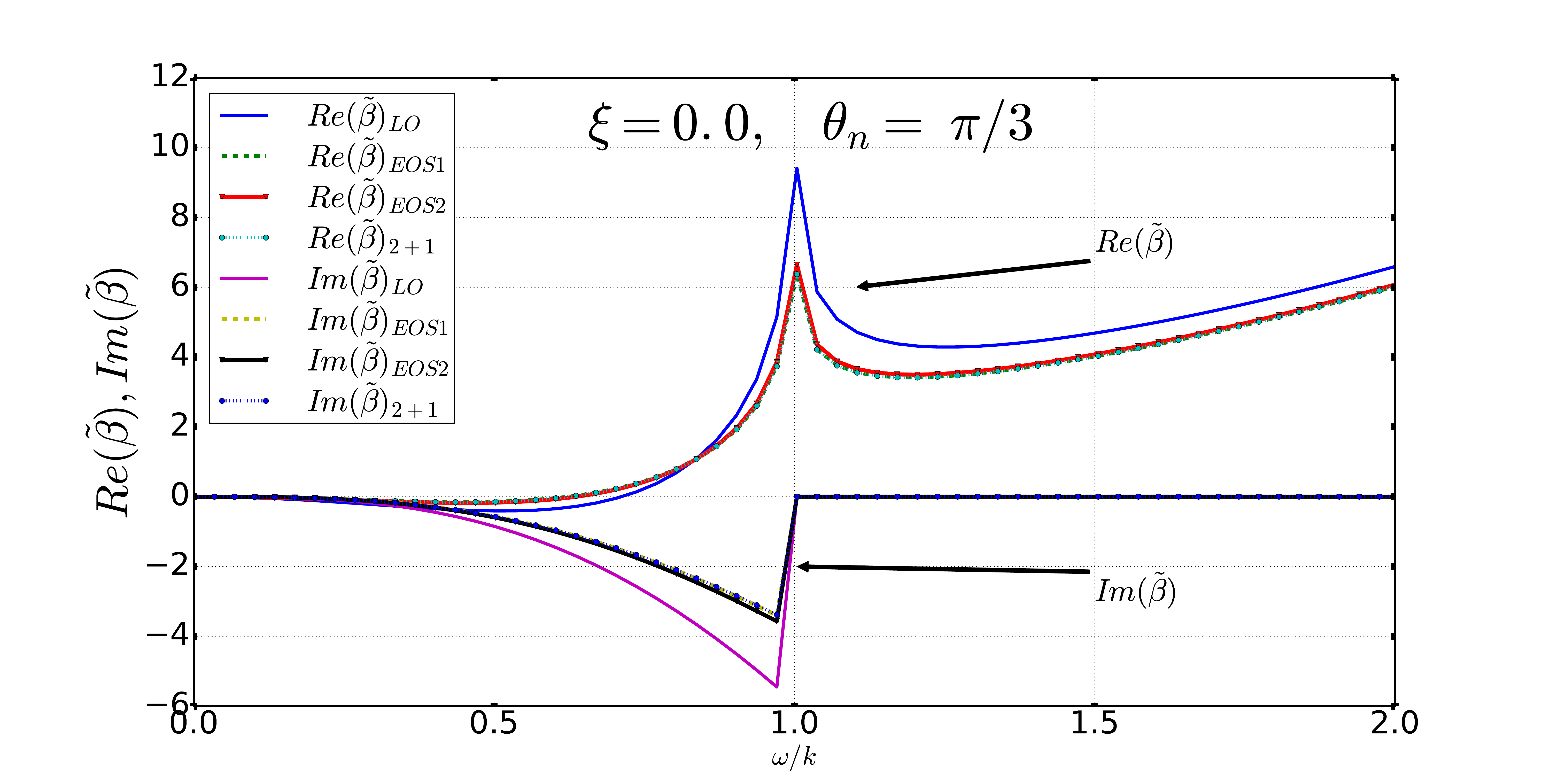}}
\subfloat{\includegraphics[height=5cm,width=6cm]{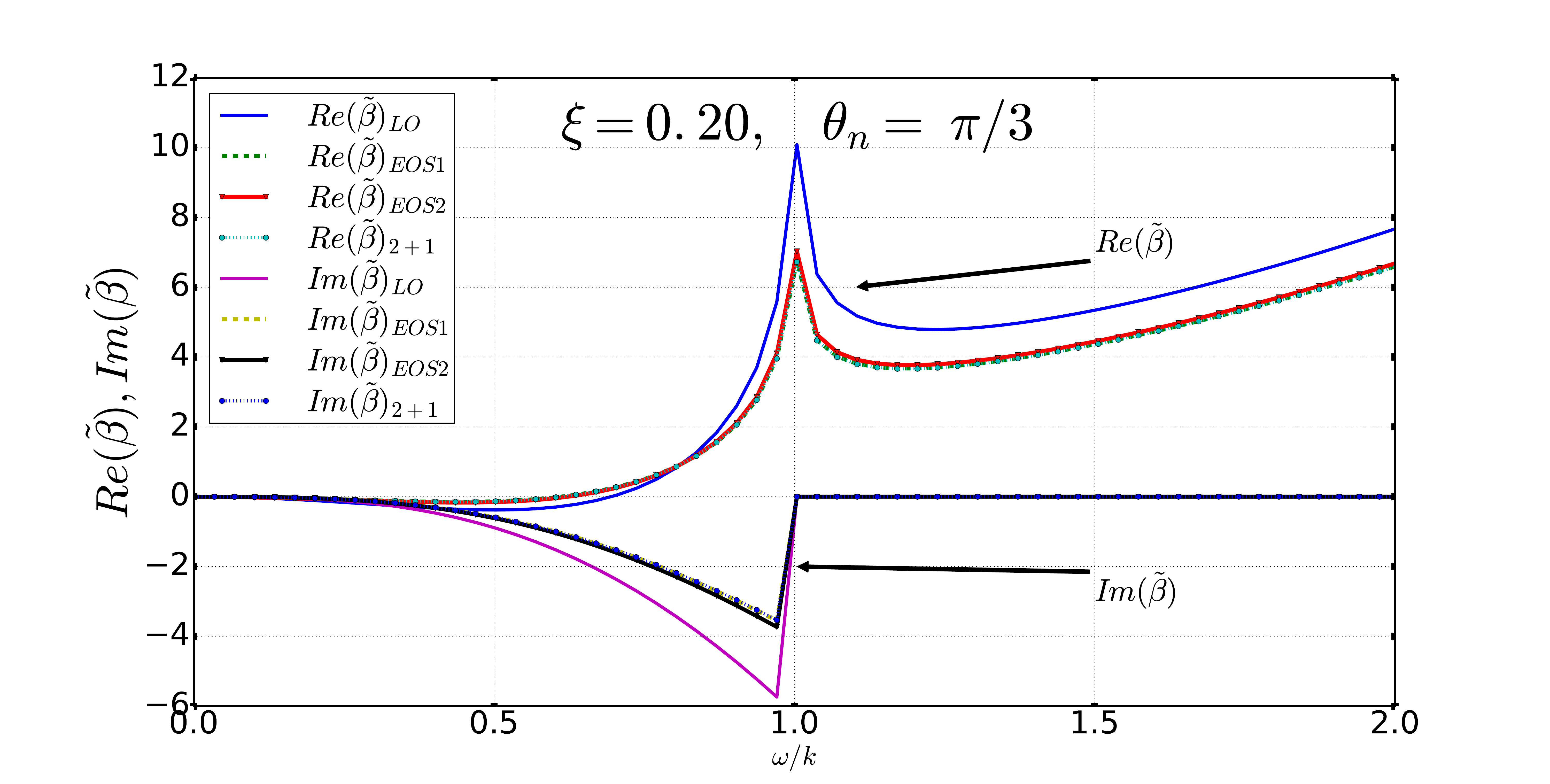}}
\subfloat{\includegraphics[height=5cm,width=6cm]{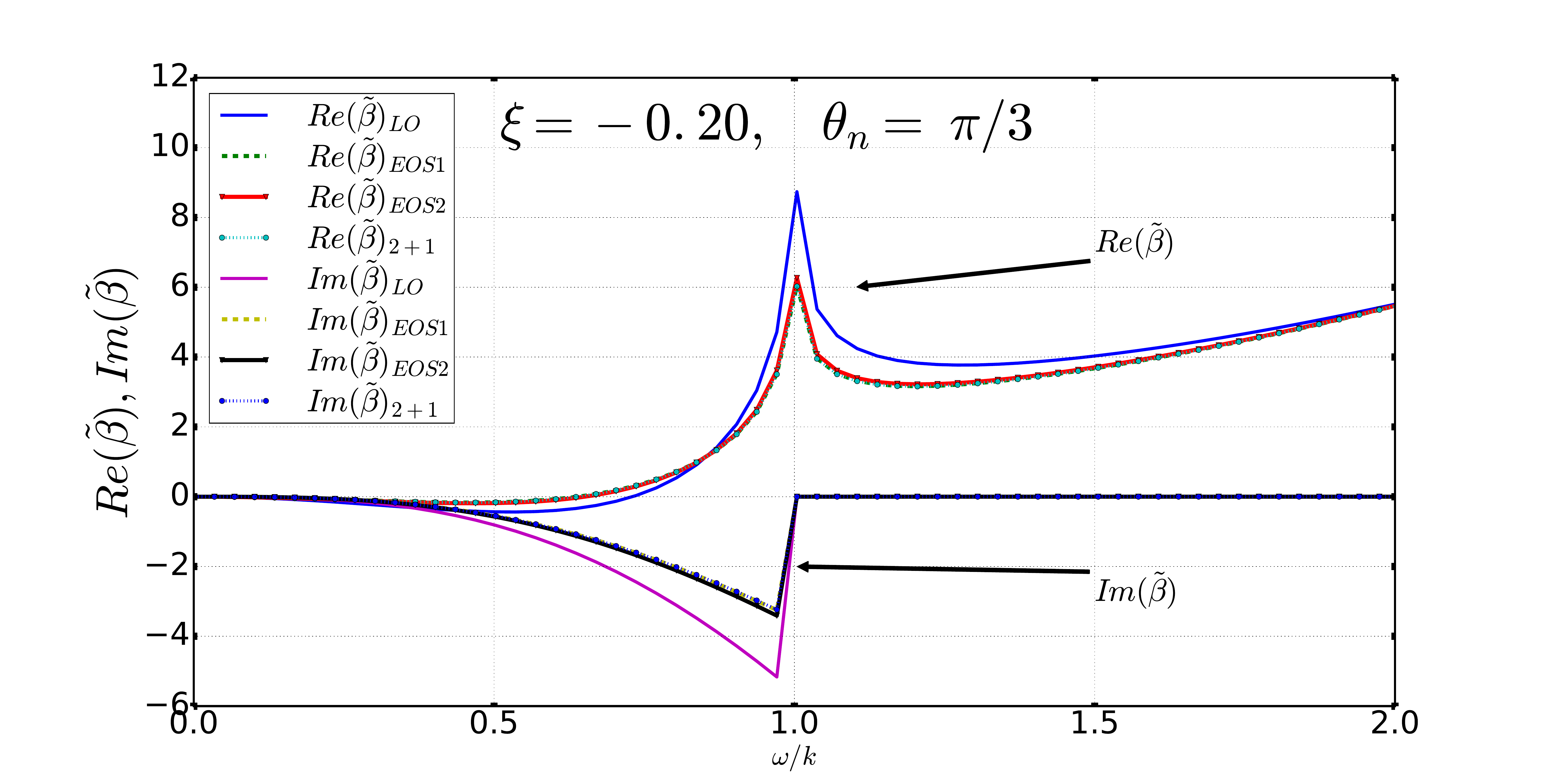}}
\caption{(Color online) $Re(\tilde{\beta})$ and $Im(\tilde{\beta})$ with $\omega/k$ at $T_{c} = 0.17GeV$ and $T = 0.25$ for various EOSs.}
\label{fig:beta}
\end{figure*}
 \begin{figure*}
  \centering
\subfloat{\includegraphics[height=7cm,width=9.4cm]{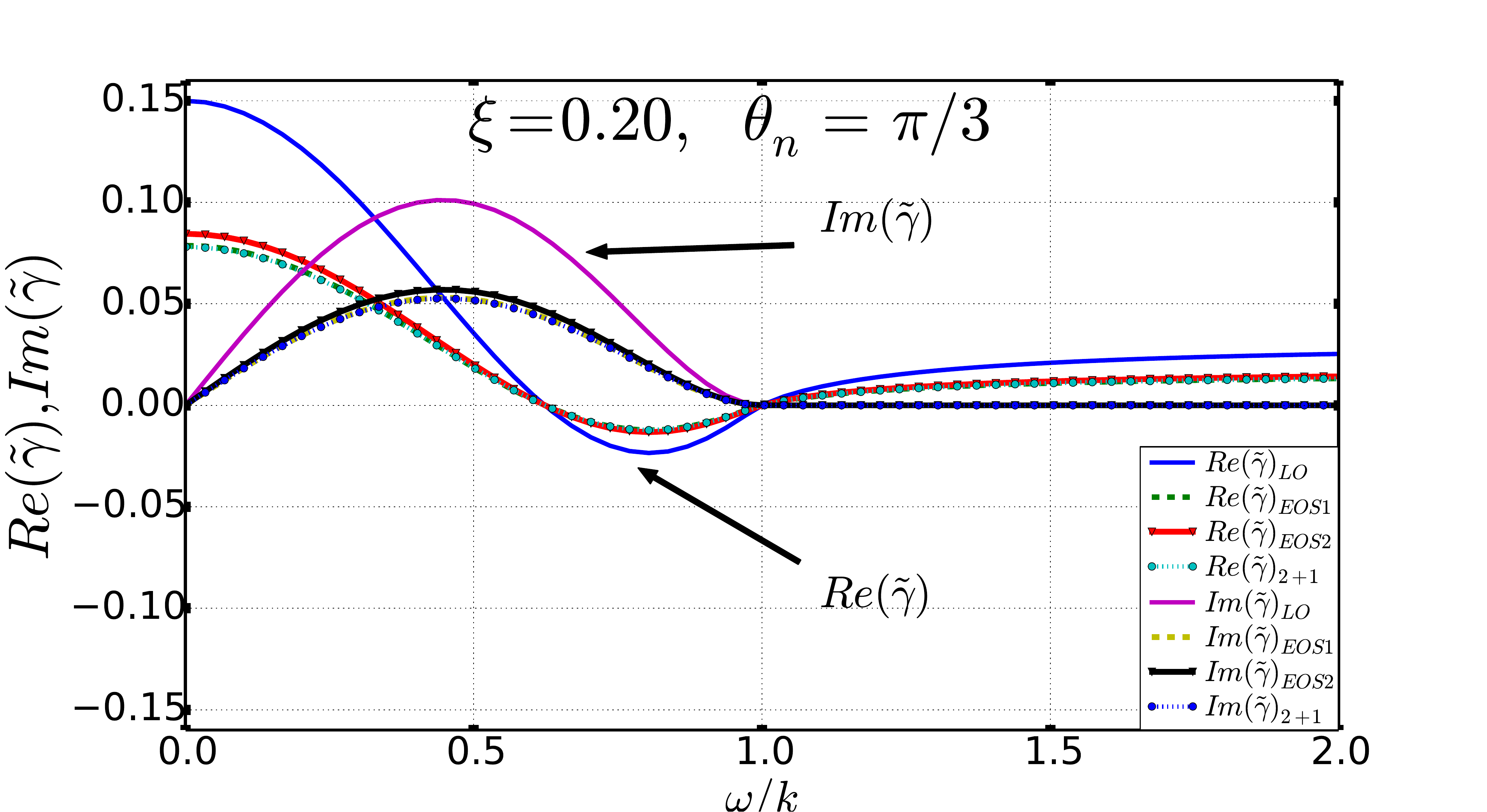}}
\subfloat{\includegraphics[height=7cm,width=9.4cm]{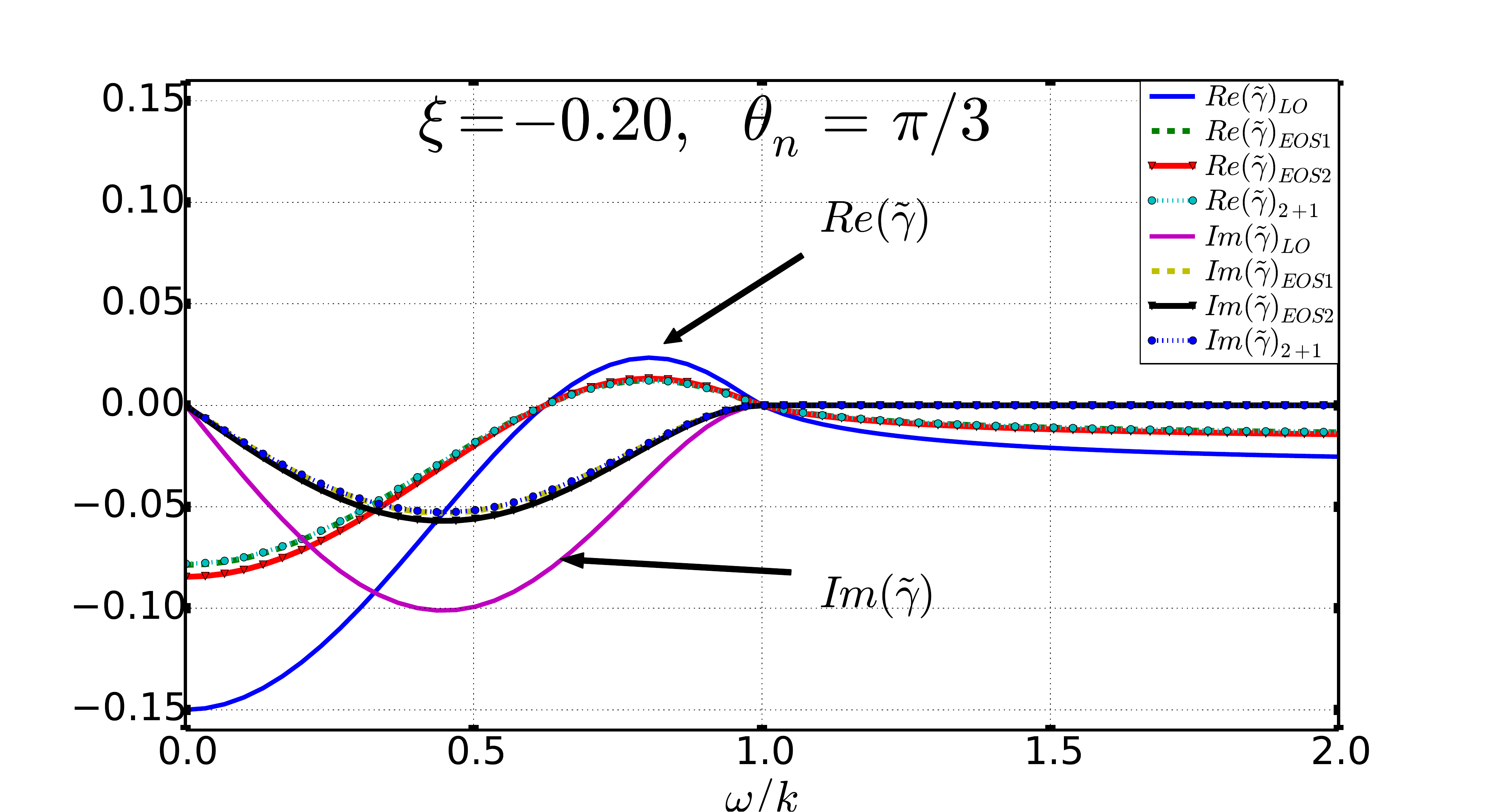}}
\caption{(Color online) $Re(\tilde{\gamma})$ and $Im(\tilde{\gamma})$ with $\omega/k$ at $T_{c} = 0.17GeV$ and $T = 0.25$ for various EOSs.}
\label{fig:gamma}
\end{figure*}

In the limit of small anisotropy, it is possible to obtain analytic expressions for all  the
four structure functions order by order in $\xi$. At the linear order in $\xi$, the distribution function 
shown in Eq.(\ref{aniso_distr}) can be expanded as,

\begin{equation}
 f_{\xi}({\bf p}) = (1 + \frac{\xi}{3}) f(p) + \frac{\xi}{2} \frac{d f(p)}{dp}({\bf p}.{\bf\hat{n}})^{2}.
 \label{dist_small_xi}
\end{equation}

Now using Eq.(\ref{pi_ij_aniso}) along with Eq.(\ref{eqpan}) and Eq.(\ref{dist_small_xi}), we can analytically obtain
the expressions for the coefficients $\alpha$, $\beta$, $\gamma$ and $\delta$. All the four coefficients have contributions of order $\xi$.
But in our analysis, we will show in the next section that $\delta$ appears quadratically in the dispersion Eq.(\ref{mode_g}), in the
linear order approximation we can neglect it. The remaining coefficients are given as,
\ba
\alpha(\omega ,{\bf k})&=&(1+\frac{\xi}{3})\alpha_{iso}(\omega ,{\bf k}) -\xi\frac{m_D^{2}(T)}{8} \bigg[\frac{8}{3}\cos^2\theta_n \nn&&
+\frac{2}{3}(5-19\cos^2\theta_n)\frac{\omega^2}{k^2}-2(1 - 5\cos^2\theta_n)\frac{\omega^4}{k^4}\nn&&+\Big[1 - 3\cos^2\theta_n 
- (2-8\cos^2\theta_n)\frac{\omega^2}{k^2} + (1 \nn&&- 5\cos^2\theta_n)\frac{\omega^4}{k^4}\Big]\frac{\omega}{k}\log \Big({\omega+k + i0^+\over\omega-k+i0^+}\Big) \bigg]
\ea 
 
\ba
\beta(\omega ,{\bf k})&=&(1+\frac{\xi}{3})\beta_{iso}(\omega ,{\bf k}) -\xi\ m_D^{2}(T) \bigg[(-\frac{2}{3}\nn&&
+\cos^2\theta_n)\frac{\omega^2}{k^2}+(1-3\cos^2\theta_n)\frac{\omega^4}{k^4}\nn&&
+\frac{1}{2}\Big[(1-2\cos^2\theta_n)\frac{\omega^2}{k^2}-(1 \nn&&
- 3\cos^2\theta_n)\frac{\omega^4}{k^4}\Big]\frac{\omega}{k}\log\Big({\omega+k + i0^+\over\omega-k+i0^+}\Big) \bigg]\nn
\ea 
  
\ba
\gamma(\omega ,{\bf k})&=&-\xi\frac{m_D^{2}(T)}{4}\sin^2\theta_n\bigg[-\frac{4}{3}+\frac{10\omega^2}{3k^2}-2\frac{\omega^4}{k^4}\nn&&
+\Big(1 - 2\frac{\omega^2}{k^2} + \frac{\omega^4}{k^4} \Big)\frac{\omega}{k}\log\Big({\omega+k + i0^+\over\omega-k+i0^+}\Big) \bigg]\nn
\ea 
where $\alpha_{iso}$ and $\beta_{iso}$ are given in Eq.(\ref{coe_iso}).

For finite $\xi$ the analytic structure of the structure functions is the same as in the isotropic case.
For real valued $\omega$ the structure functions are real for all $\omega > k$ and complex for $\omega < k$.
For imaginary values of $\omega$ all four structure functions are real.
Since the dispersion relation can't be solved analytically for arbitrary $k$, we are considering here the limit when $\omega \gg k$.
In this limit the coefficients $\alpha$, $\beta$ and $\gamma$ can be obtained as,
\ba
\alpha(\omega ,{\bf k})&=&m_D^{2}(T)\bigg[\frac{1}{3}(1-\frac{\xi}{15})+\frac{1}{5} \Big[\frac{1}{3}\nn&&
+ \frac{\xi}{7}(\frac{1}{9} + \cos^2\theta_n)\Big]\frac{\omega^2}{k^2}+ O\bigg(\frac{\omega^4}{k^4}\bigg)\bigg]
\ea 

\ba
\beta(\omega ,{\bf k})&=&m_D^{2}(T)\bigg[\frac{1}{3}\Big[1+\frac{\xi}{5}(-\frac{1}{3}+\cos^2\theta_n)\Big]\nn&&
+\frac{1}{5}\Big[1+ \frac{\xi}{7}(\frac{1}{3} - \cos^2\theta_n)\Big]\frac{\omega^2}{k^2}+ O\bigg(\frac{\omega^4}{k^4}\bigg)\bigg]\nn
\ea 
 
\ba
\gamma(\omega ,{\bf k})&=&m_D^{2}(T)\xi\sin^2\theta_n\bigg[\frac{1}{15} -\frac{4k^2}{105\omega^2} + O\bigg(\frac{\omega^4}{k^4}\bigg)\bigg]\nn
\ea 
 Since, all the structure constants depend on the Debye mass $m^{2}_{D}$, any modification in Debye mass will modify all of the them.
With these structure functions in hand we can construct the $\Pi^{ij}(k)$ in the small $\xi$ limit. In the next section, we shall discuss the
formalism of finding the dispersion relation for the collective modes from the propagator shown in Eq.(\ref{eq:propagator})

\section{Collective modes}
 \begin{figure*}
  \centering
\subfloat{\includegraphics[height=5cm,width=6cm]{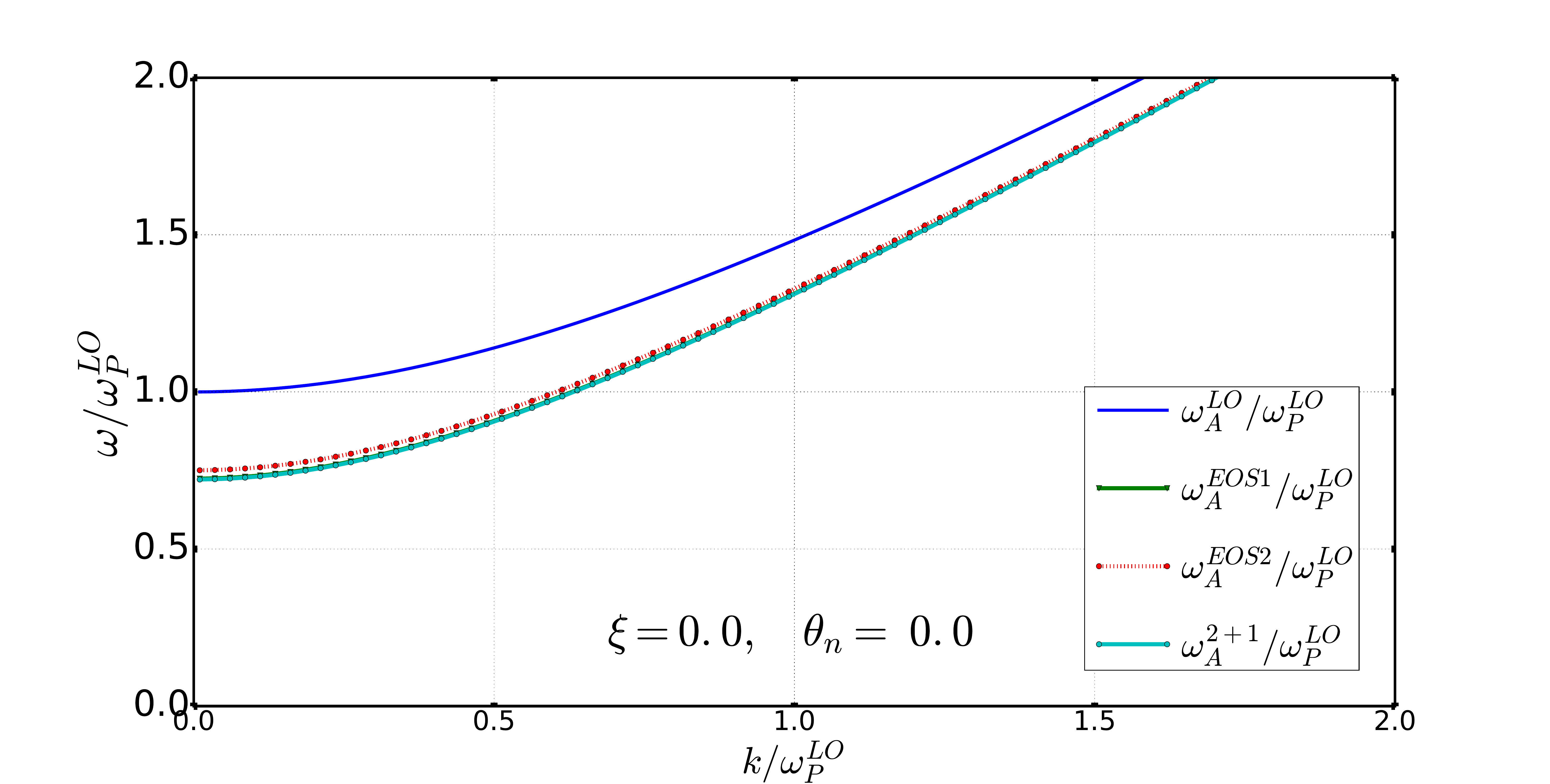}}
\subfloat{\includegraphics[height=5cm,width=6cm]{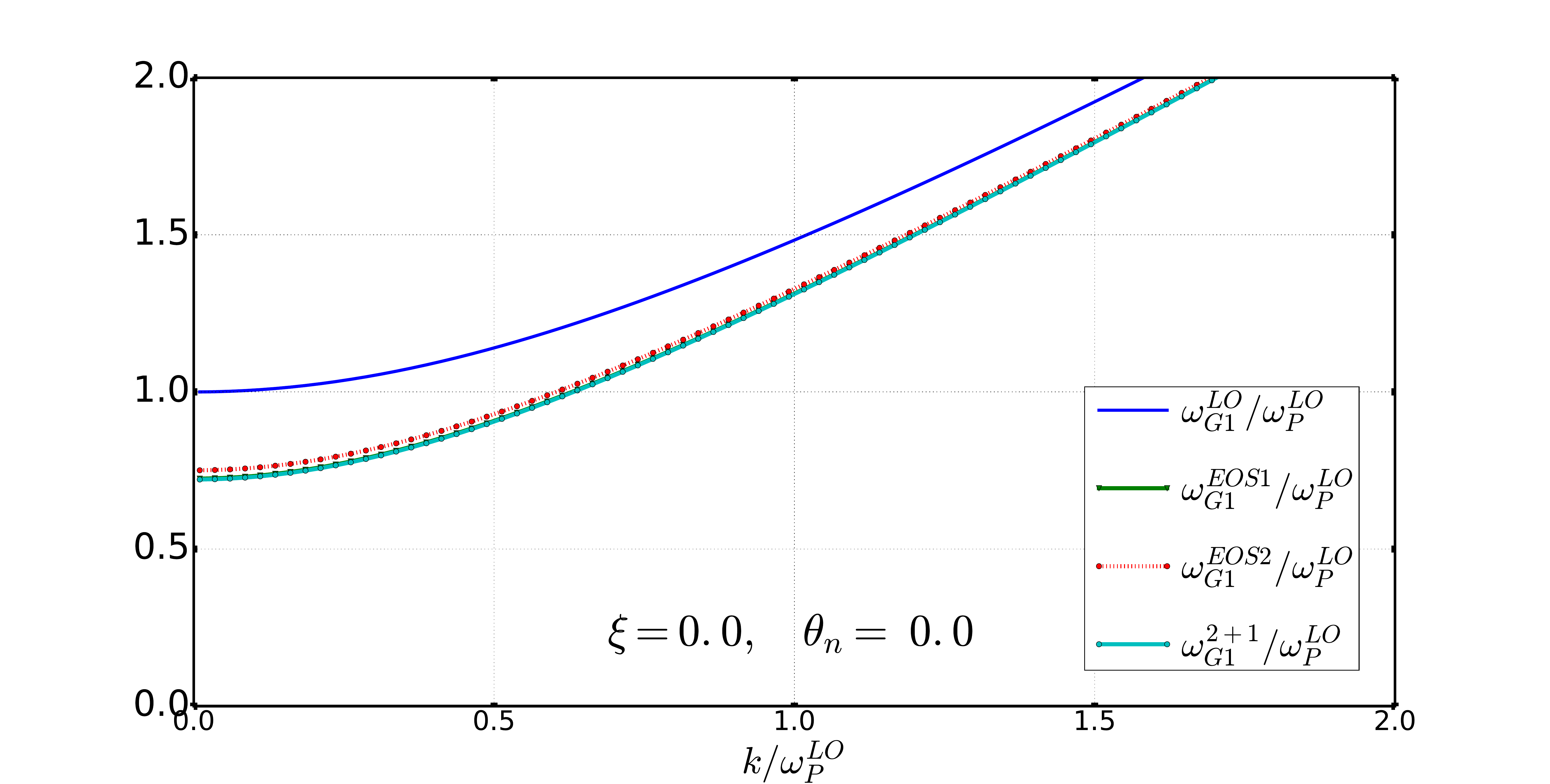}}
\subfloat{\includegraphics[height=5cm,width=6cm]{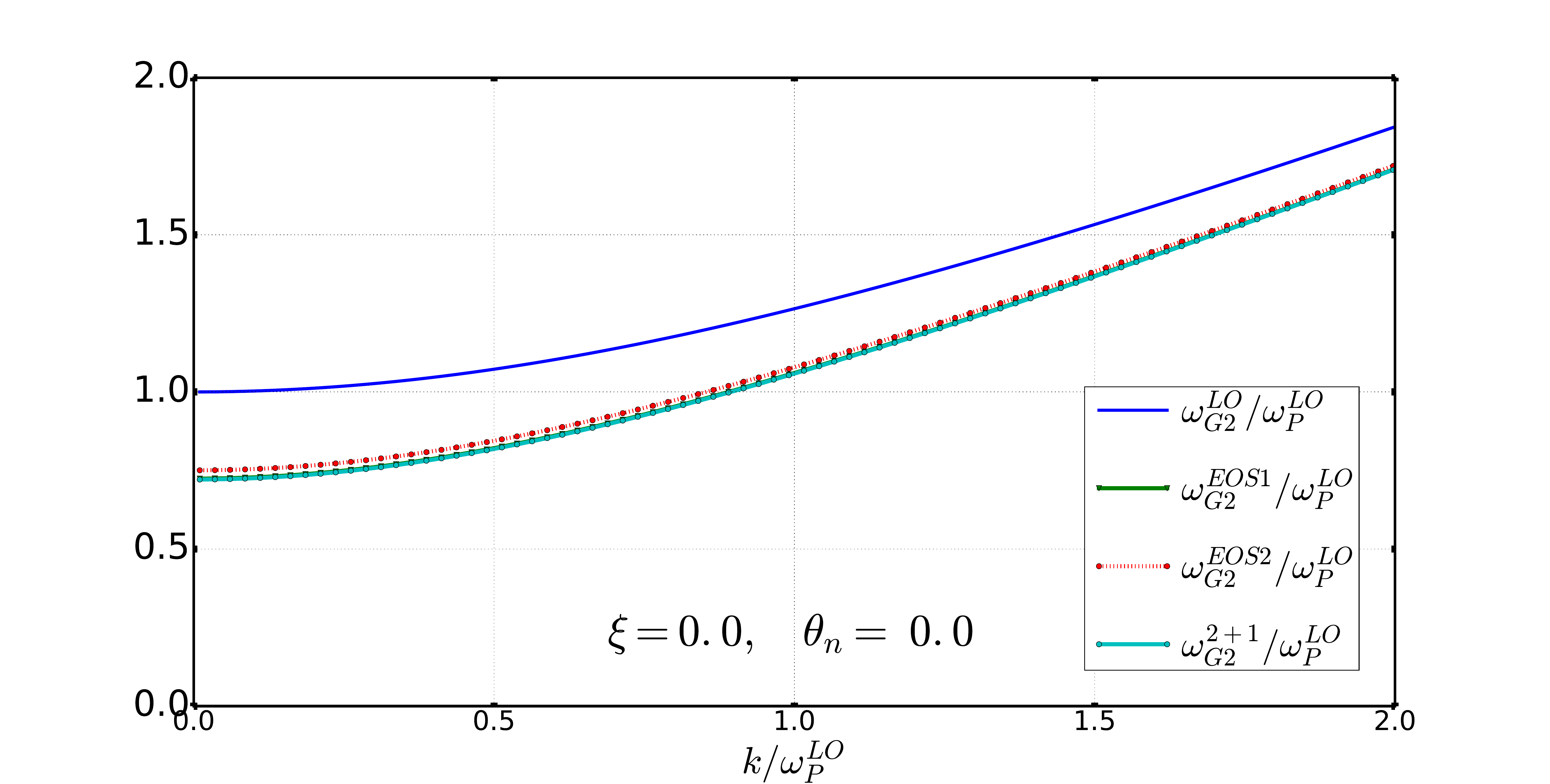}}
\vspace{3mm}
\subfloat{\includegraphics[height=5cm,width=6cm]{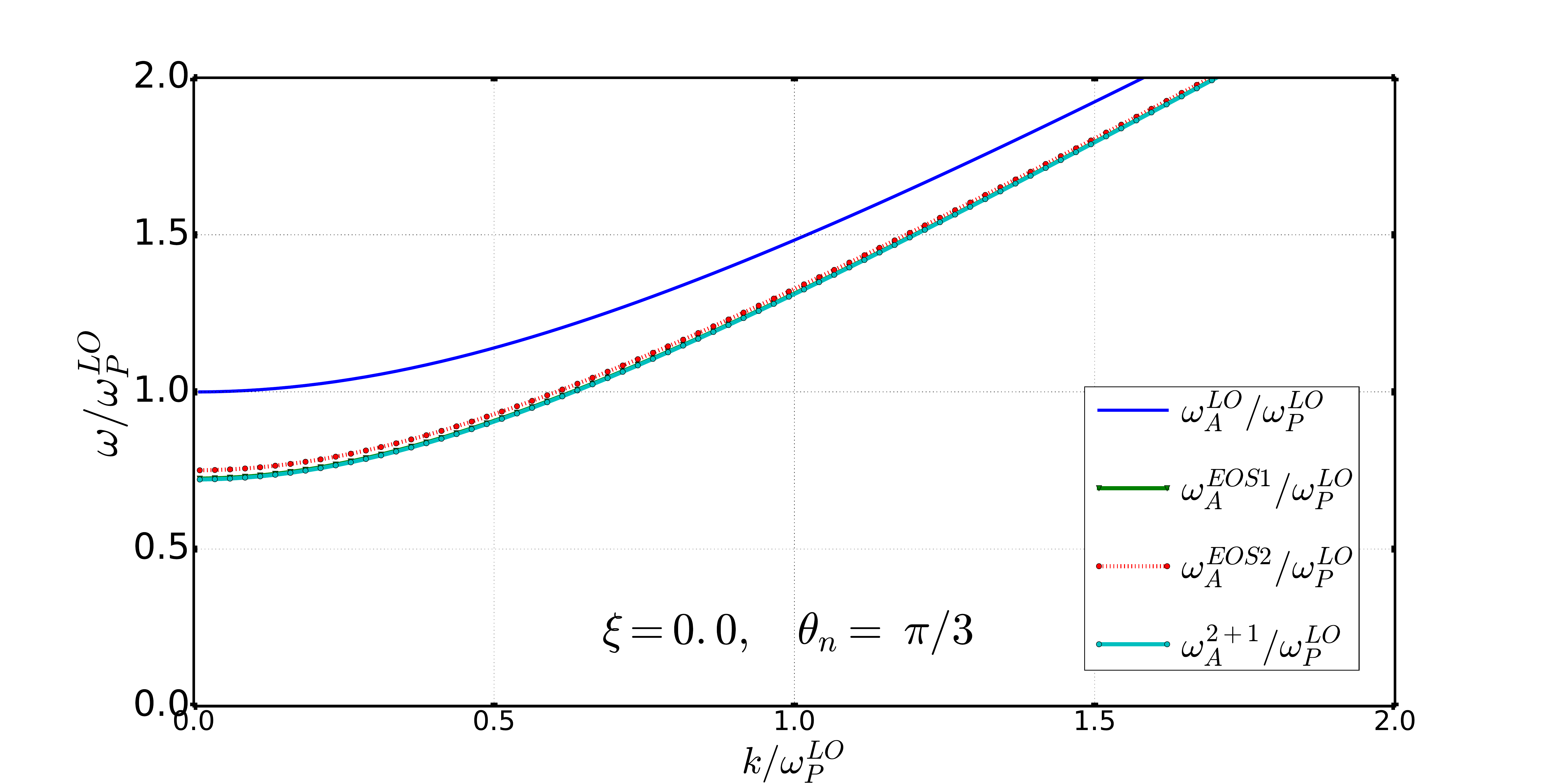}}
\subfloat{\includegraphics[height=5cm,width=6cm]{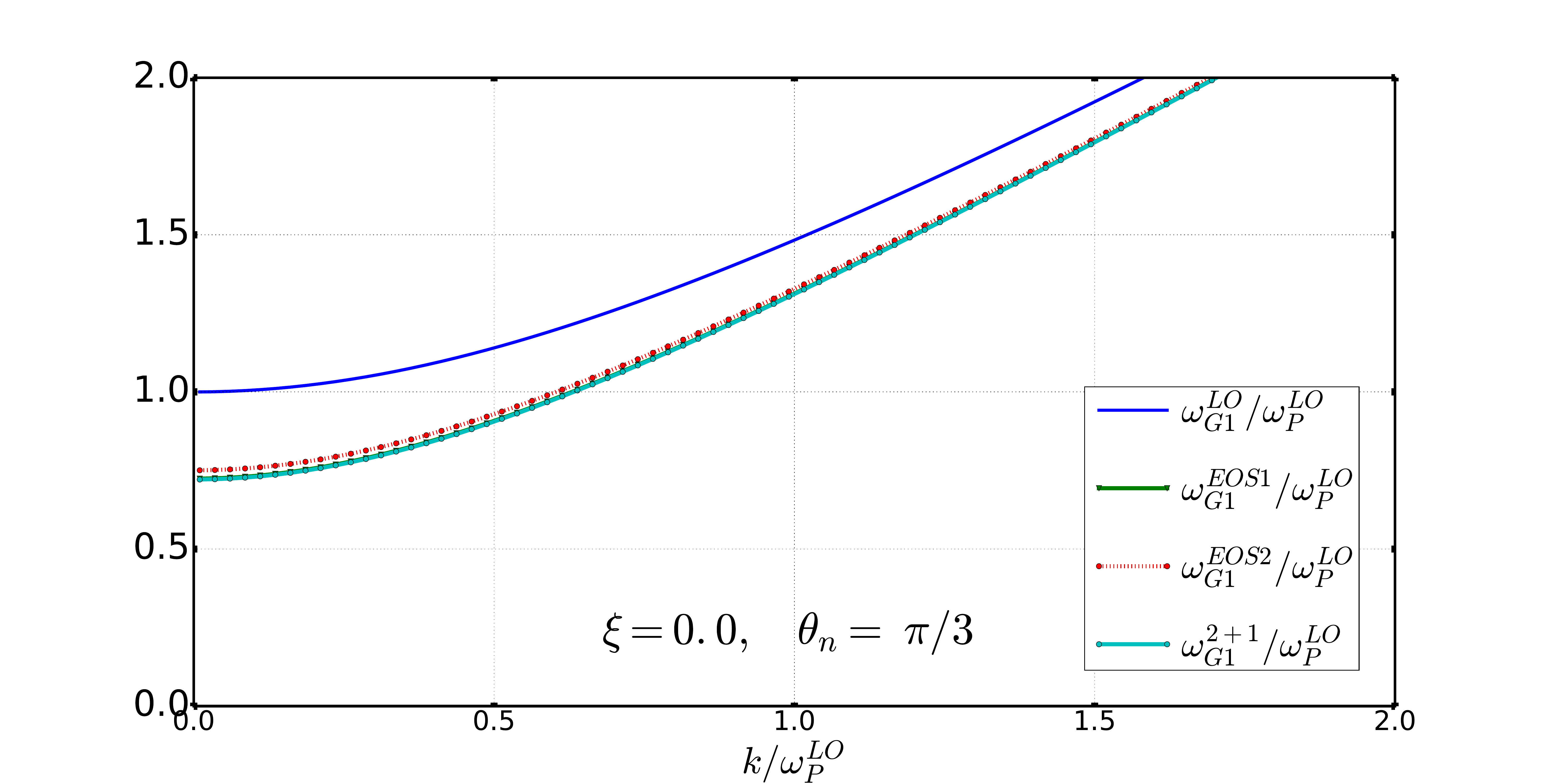}}
\subfloat{\includegraphics[height=5cm,width=6cm]{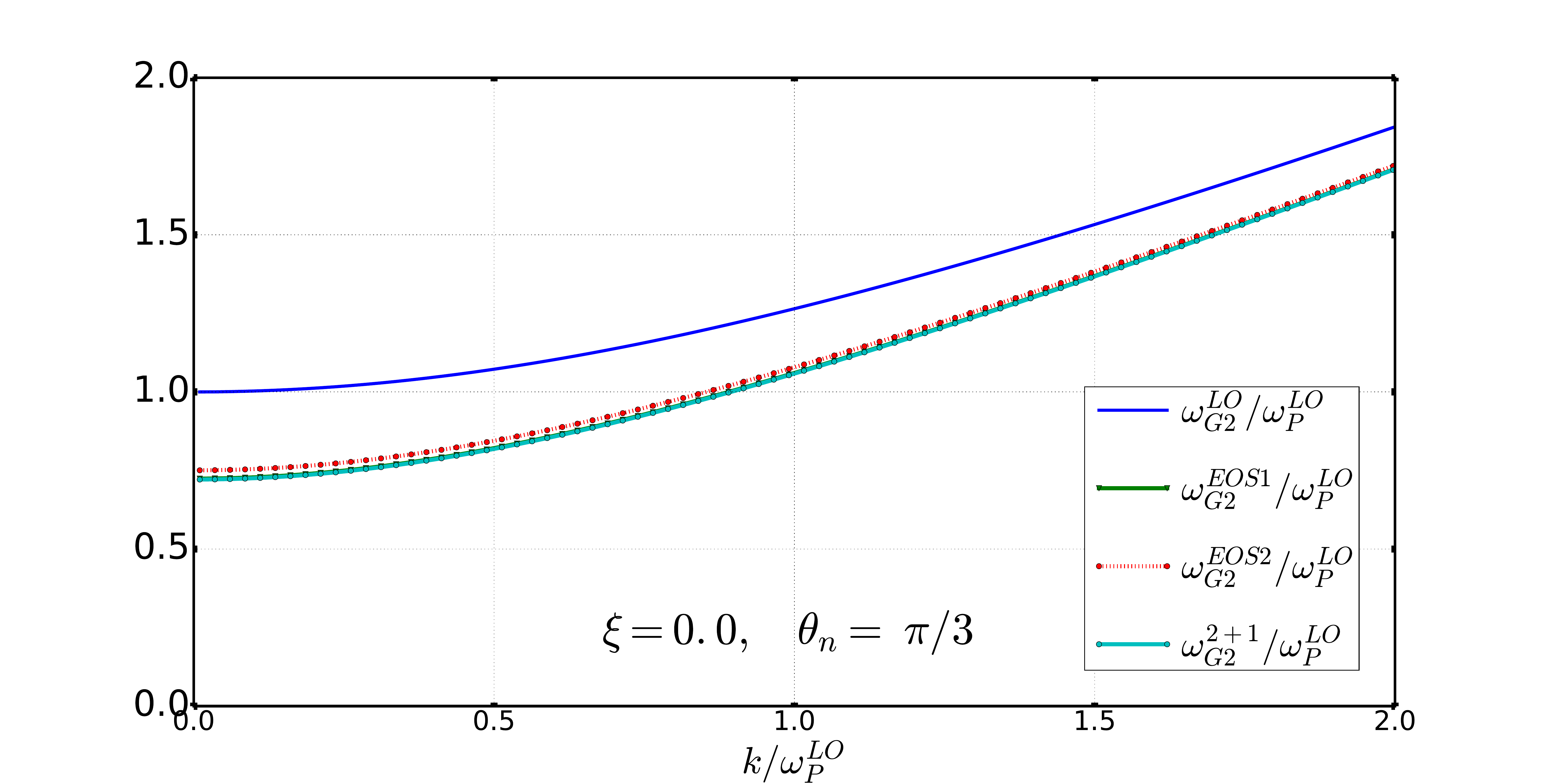}}
\caption{(color online) Dispersion curve of modes for various EOSs with different $\theta_n$ at fixed $\xi = 0.0$, $T_{c} = 0.17GeV$ and $T = 0.25GeV$.}
\label{fig:modes_0}
 \end{figure*} \begin{figure*}[]
  \centering
\subfloat{\includegraphics[height=5cm,width=6cm]{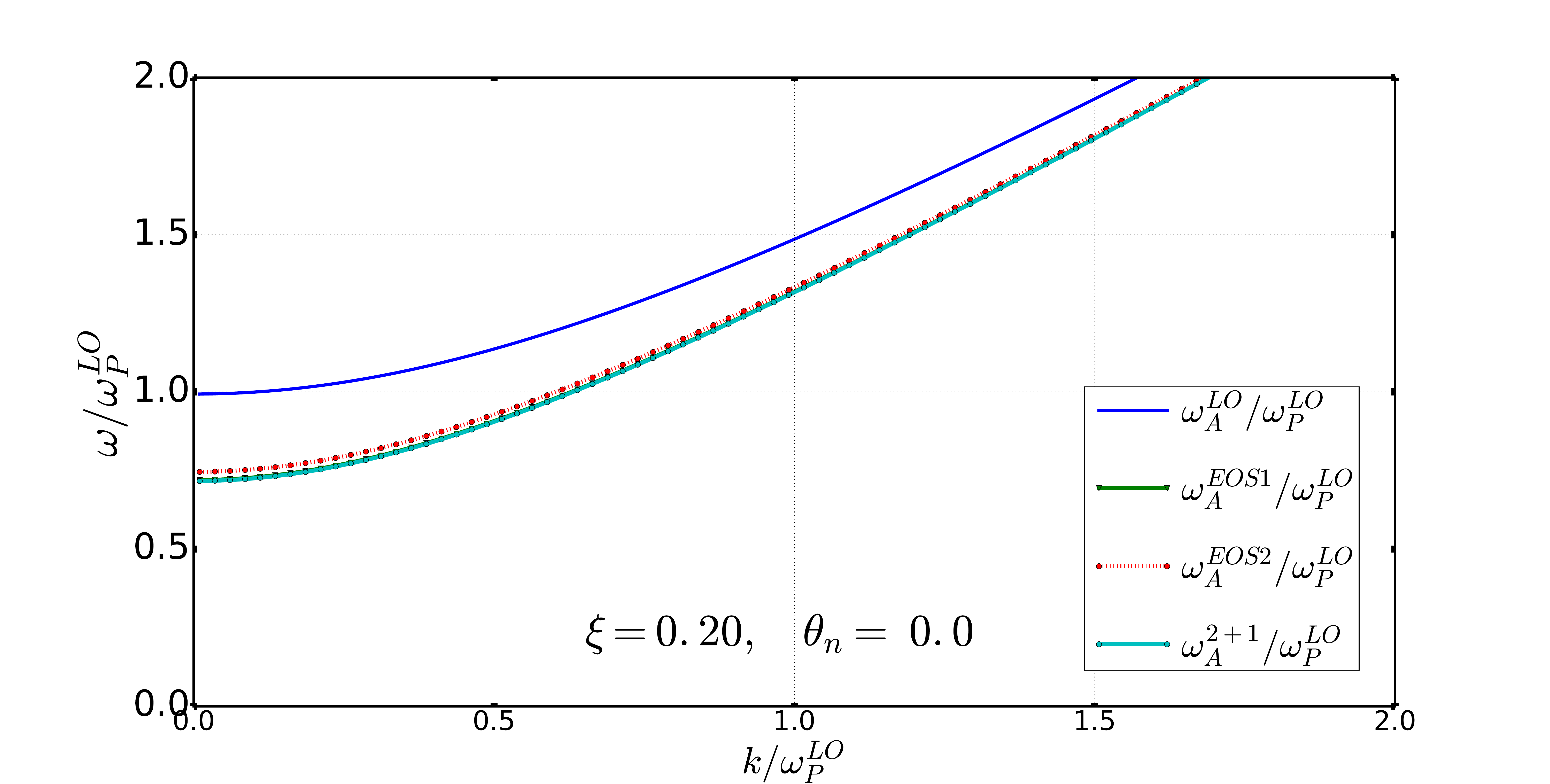}}
\subfloat{\includegraphics[height=5cm,width=6cm]{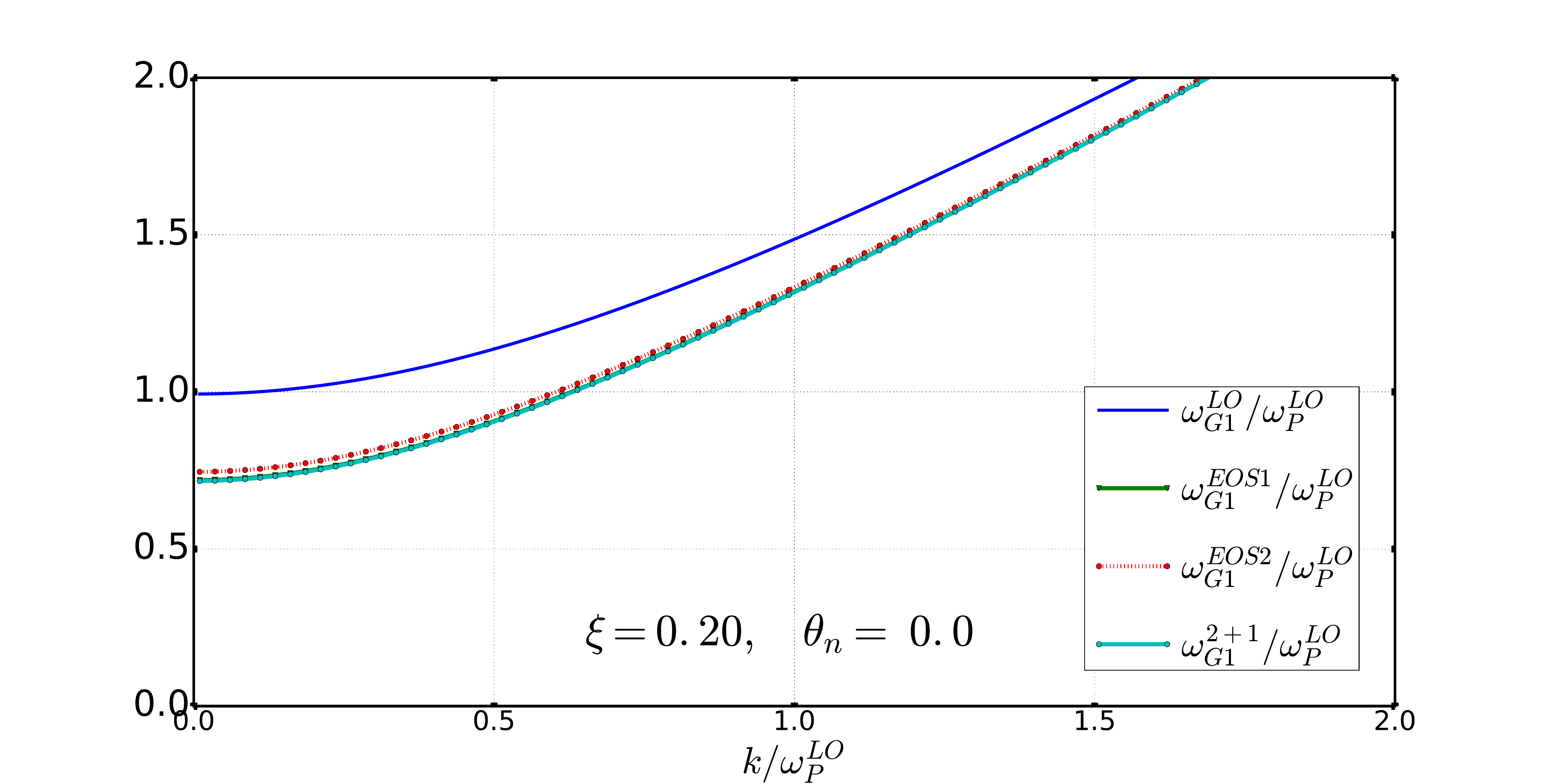}}
\subfloat{\includegraphics[height=5cm,width=6cm]{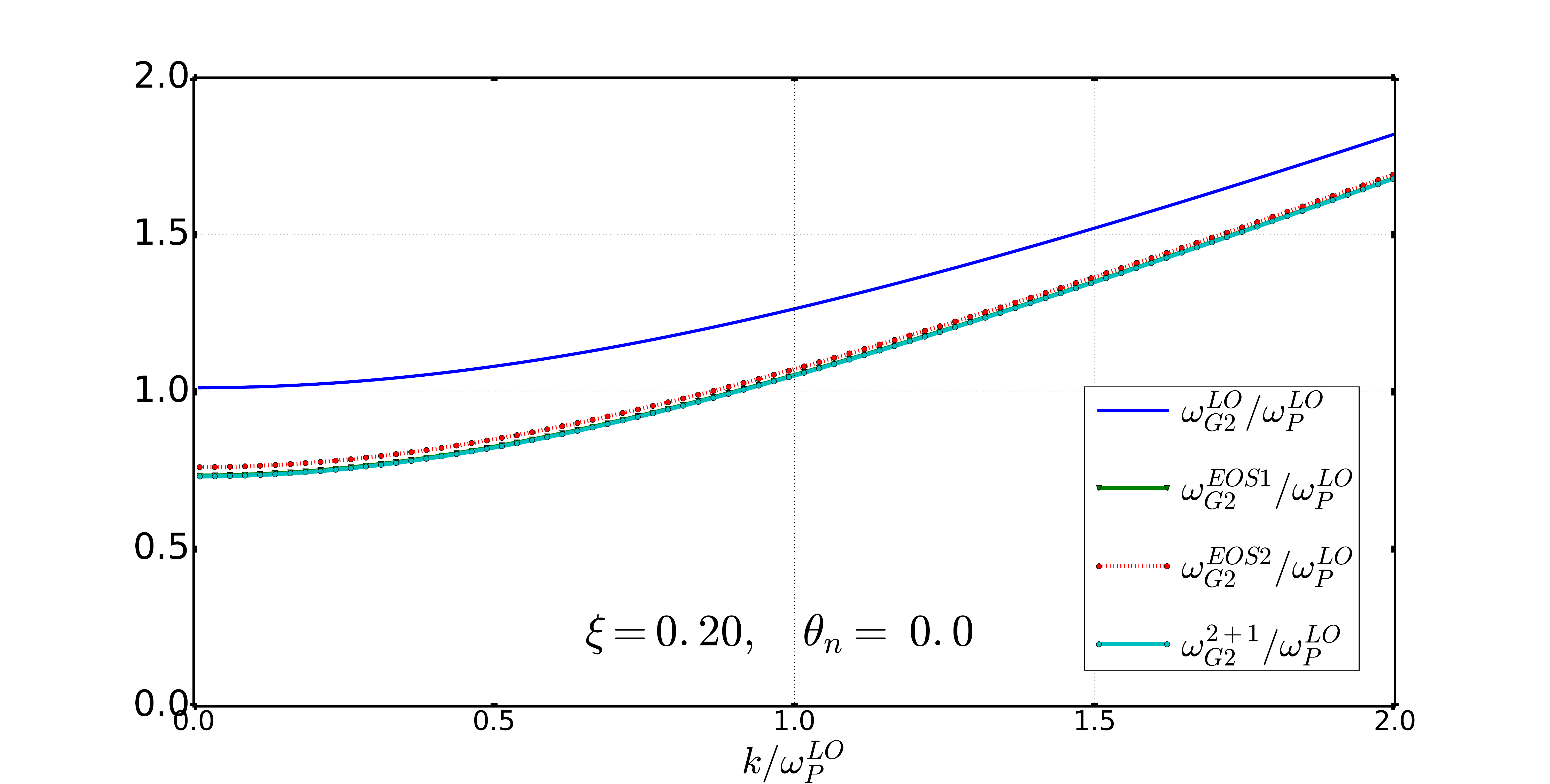}}
\hspace{3mm}
\subfloat{\includegraphics[height=5cm,width=6cm]{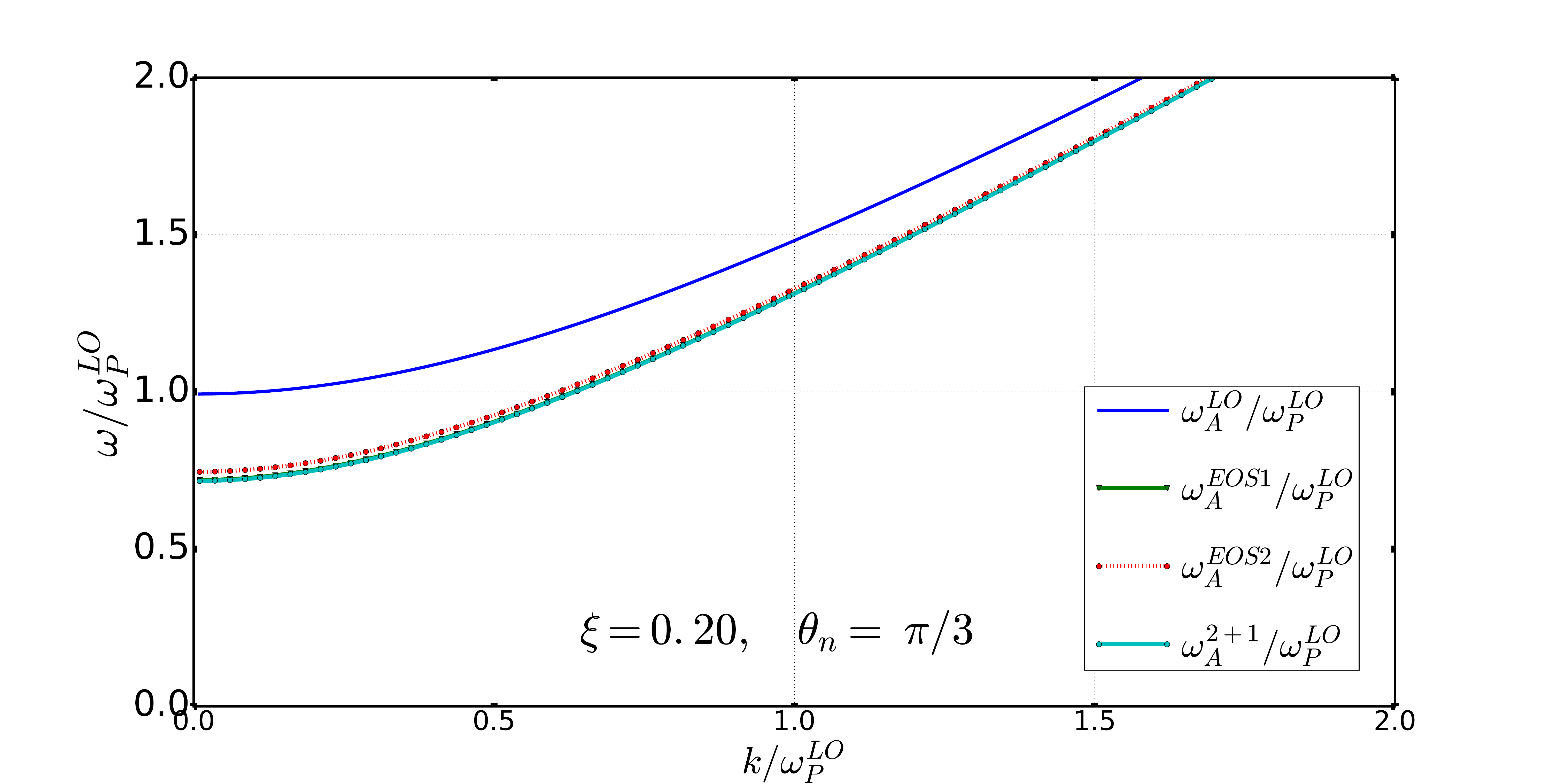}}
\subfloat{\includegraphics[height=5cm,width=6cm]{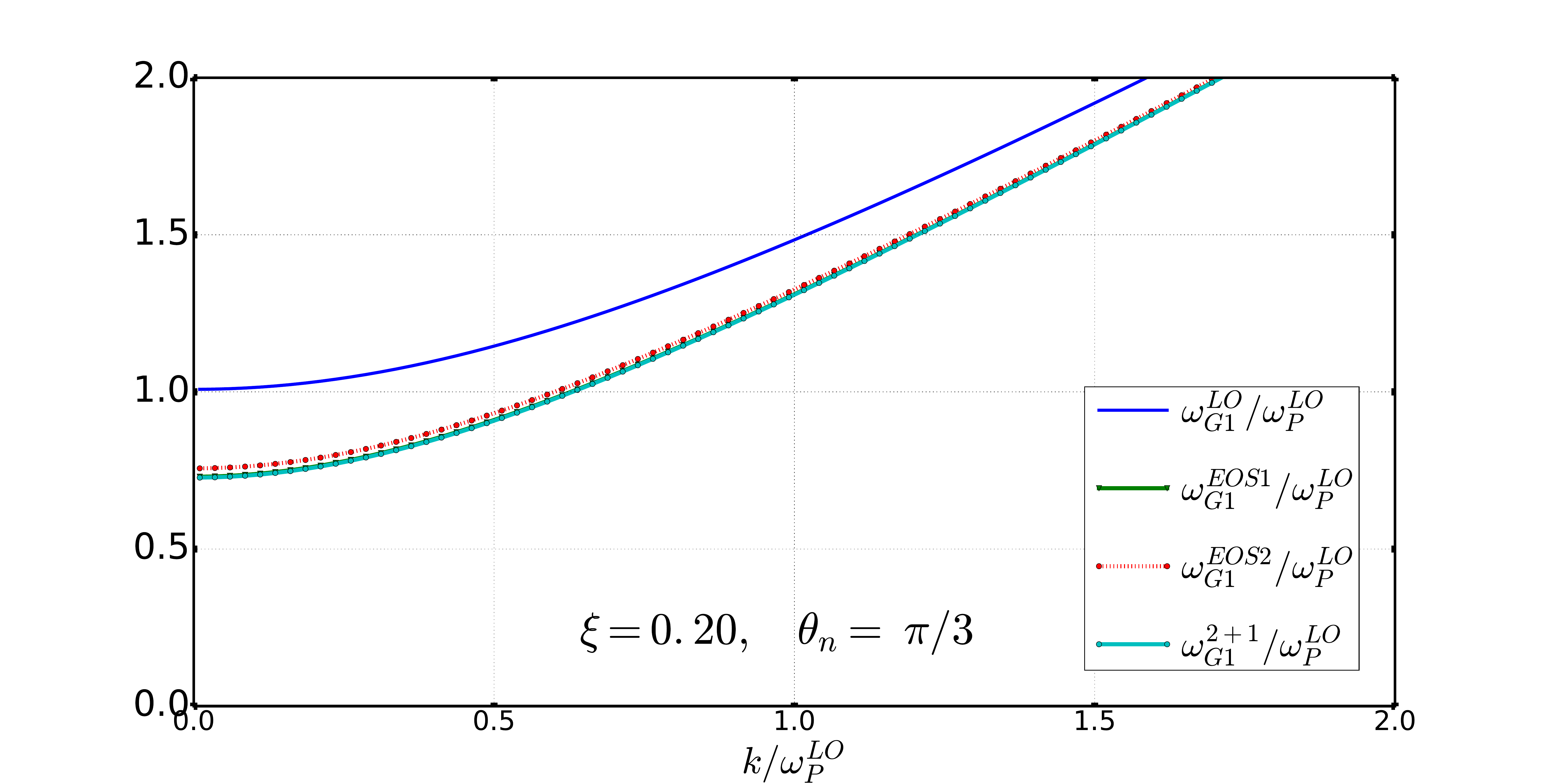}}
\subfloat{\includegraphics[height=5cm,width=6cm]{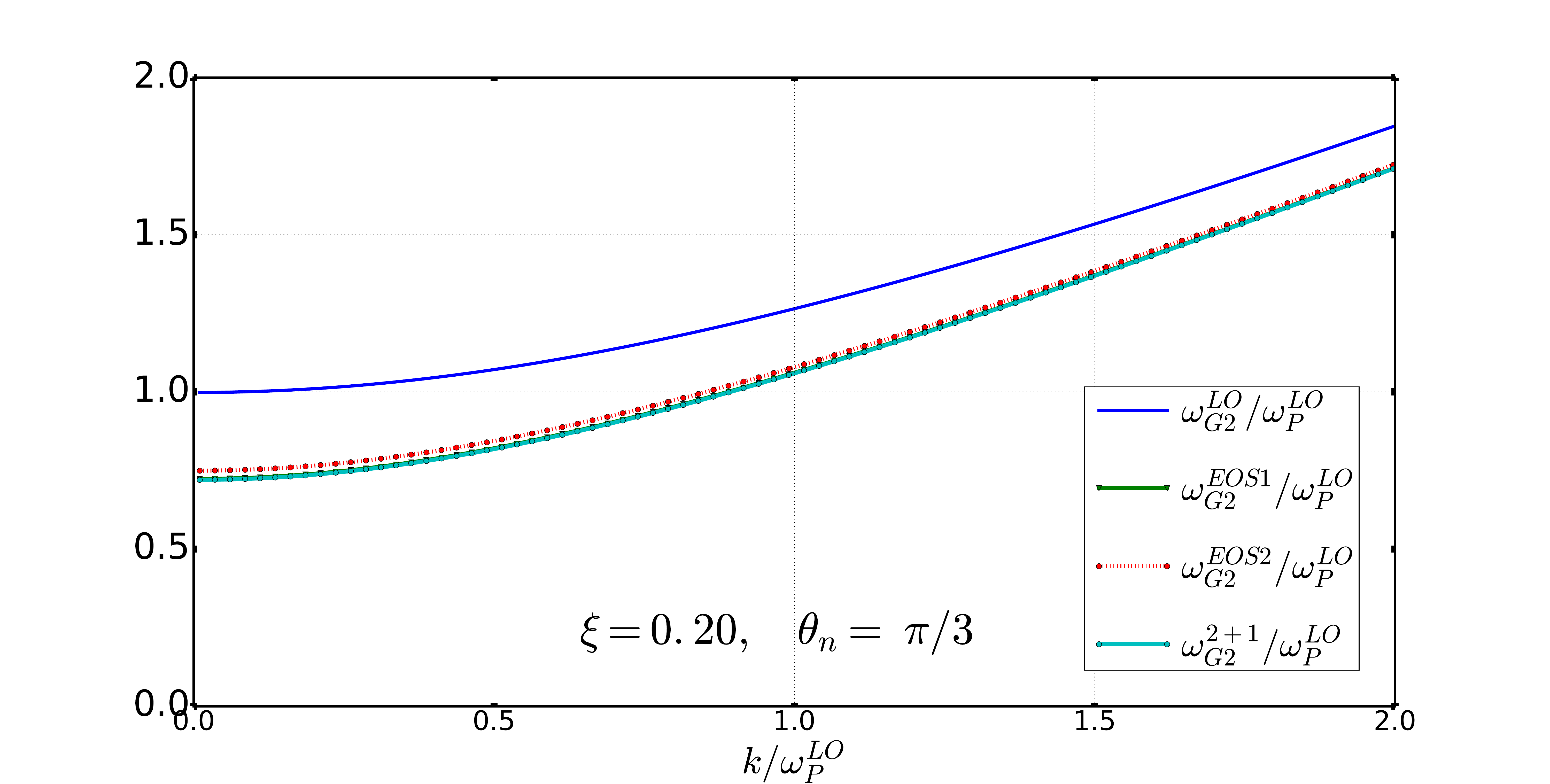}}
\caption{(color online) Dispersion curve of modes for various EOSs with different $\theta_n$ at fixed $\xi = 0.20$, $T_{c} = 0.17GeV$ and $T = 0.25GeV$.}
\label{fig:modes_2}
 \end{figure*} \begin{figure*}[]
  \centering
\subfloat{\includegraphics[height=5cm,width=6cm]{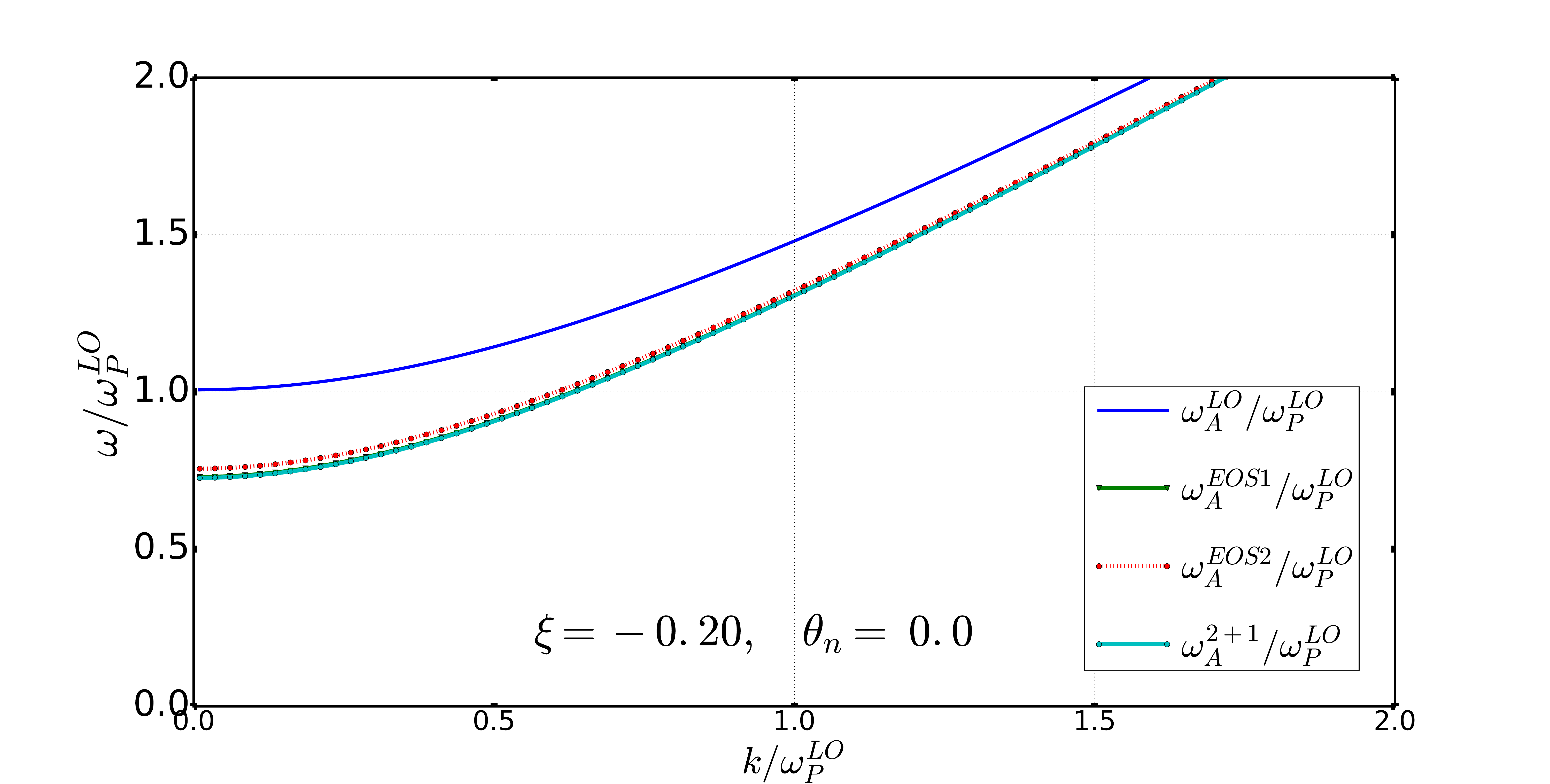}}
\subfloat{\includegraphics[height=5cm,width=6cm]{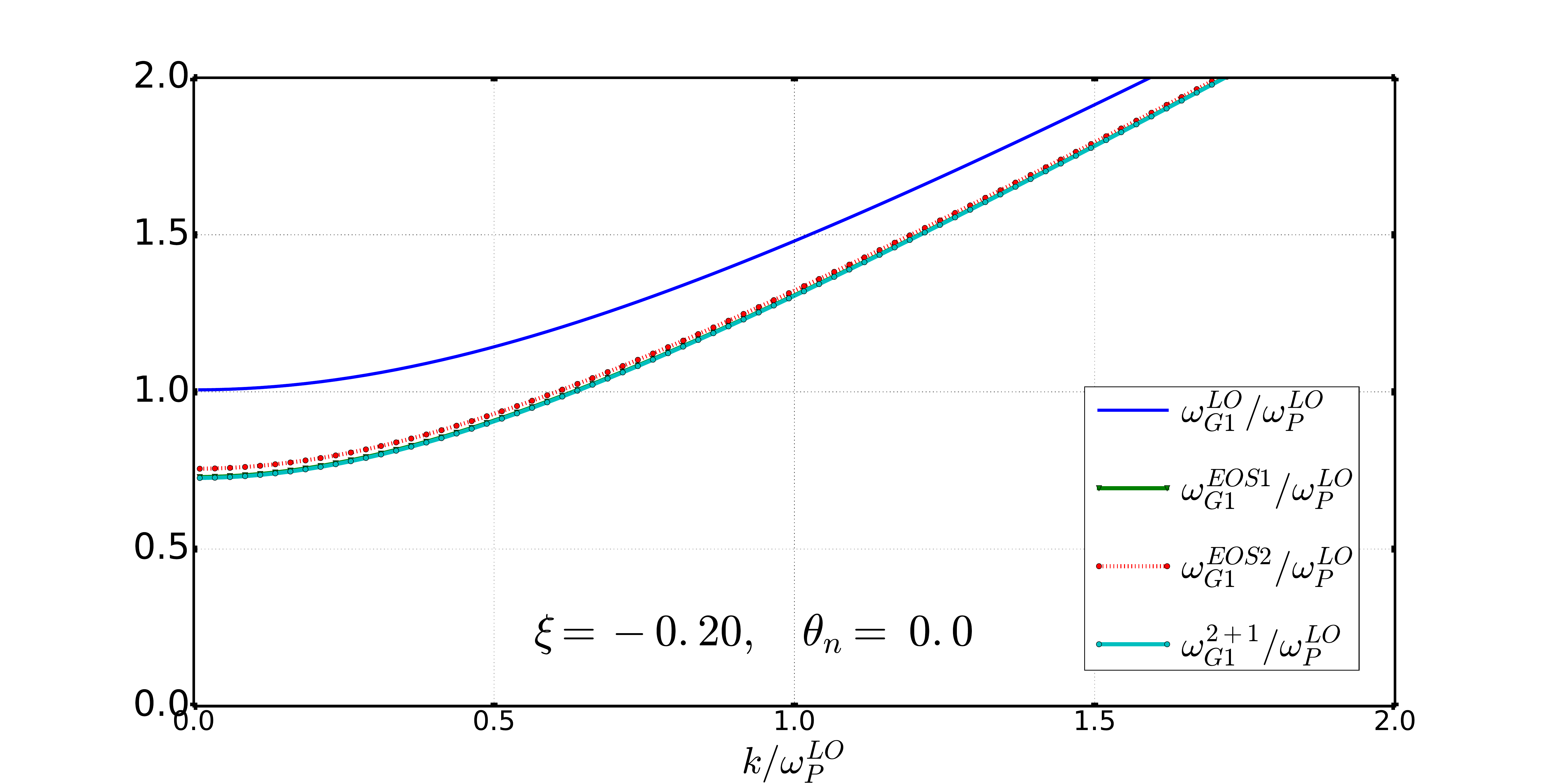}}
\subfloat{\includegraphics[height=5cm,width=6cm]{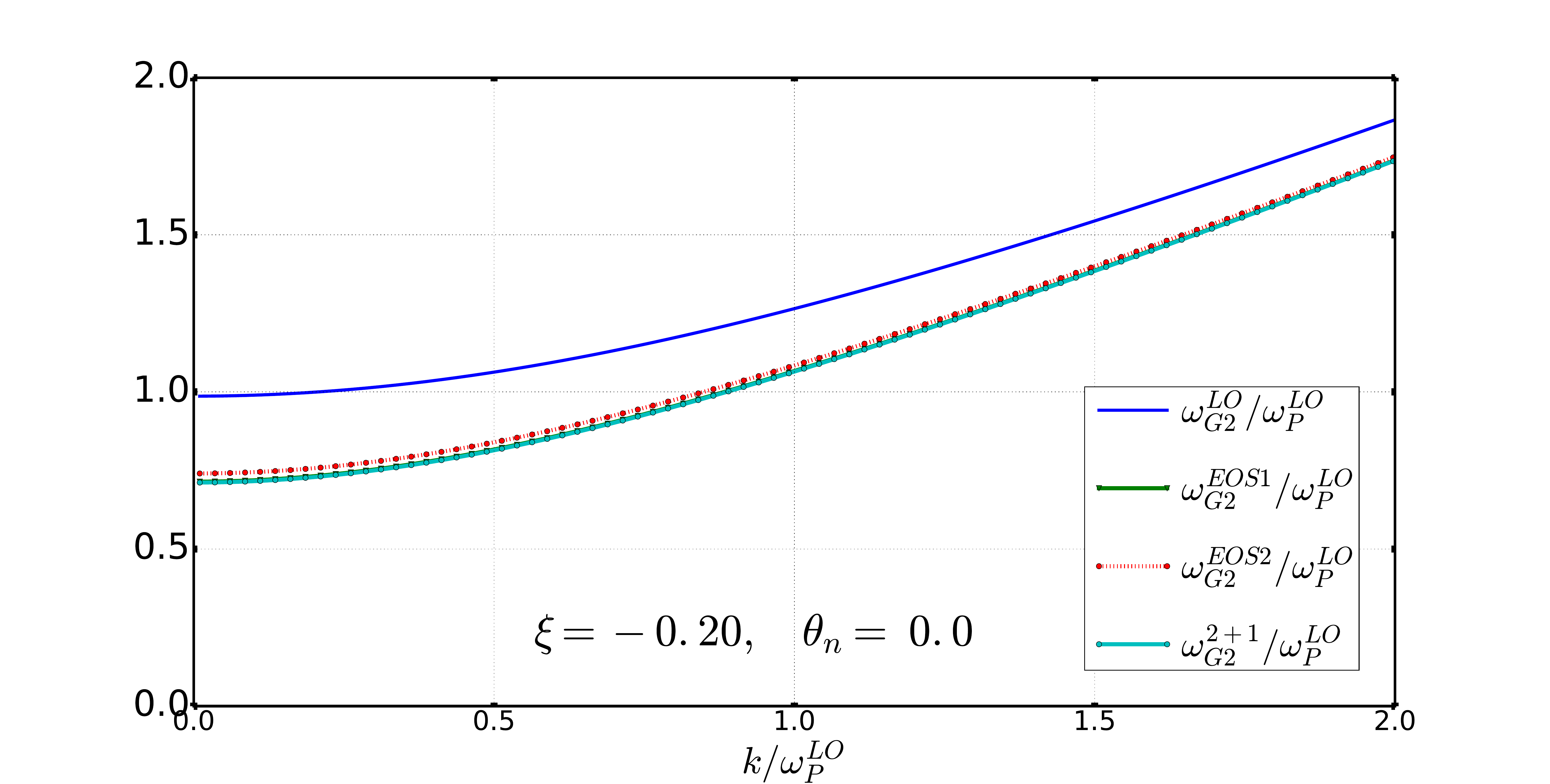}}
\hspace{3mm}
\subfloat{\includegraphics[height=5cm,width=6cm]{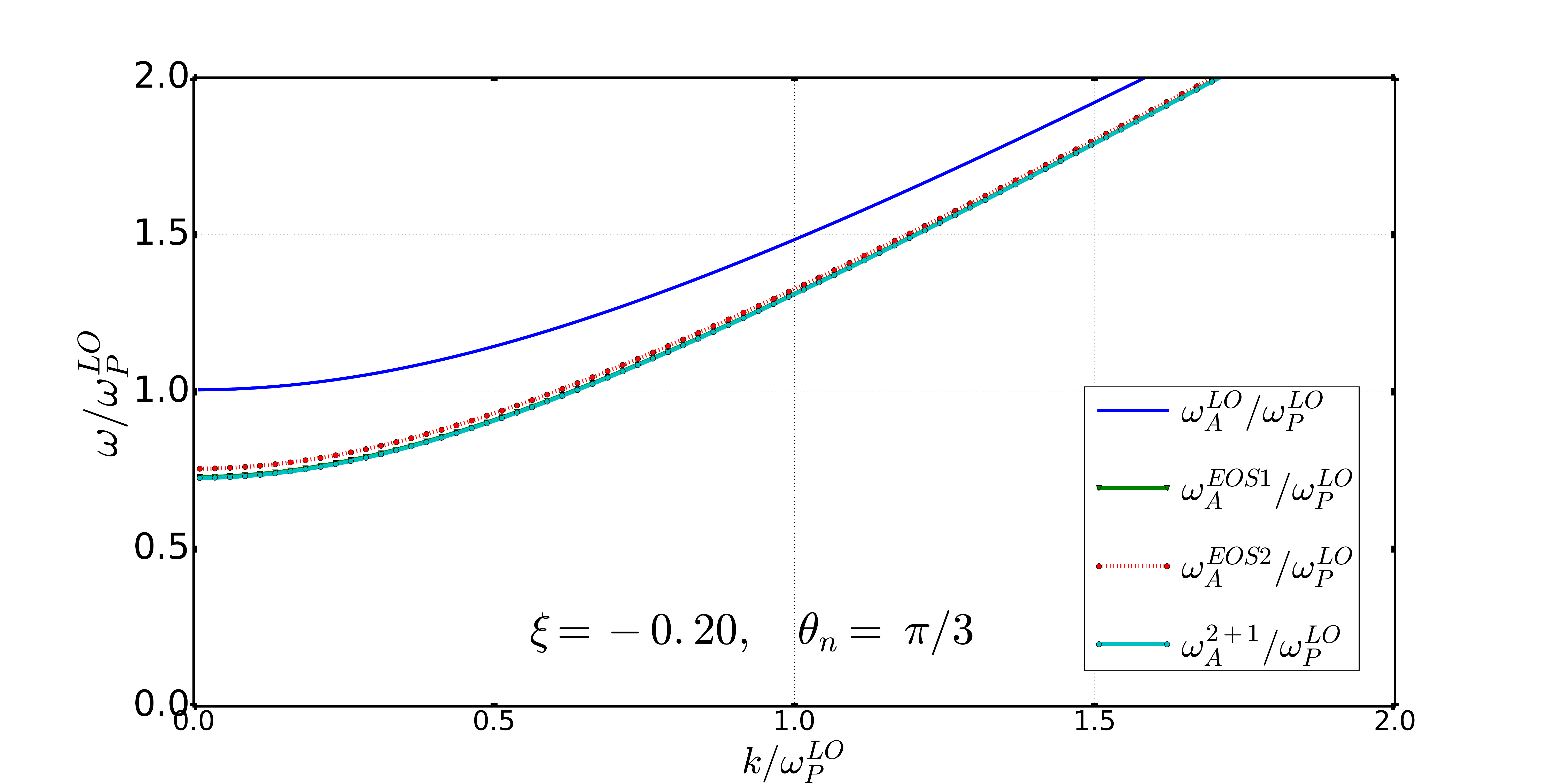}}
\subfloat{\includegraphics[height=5cm,width=6cm]{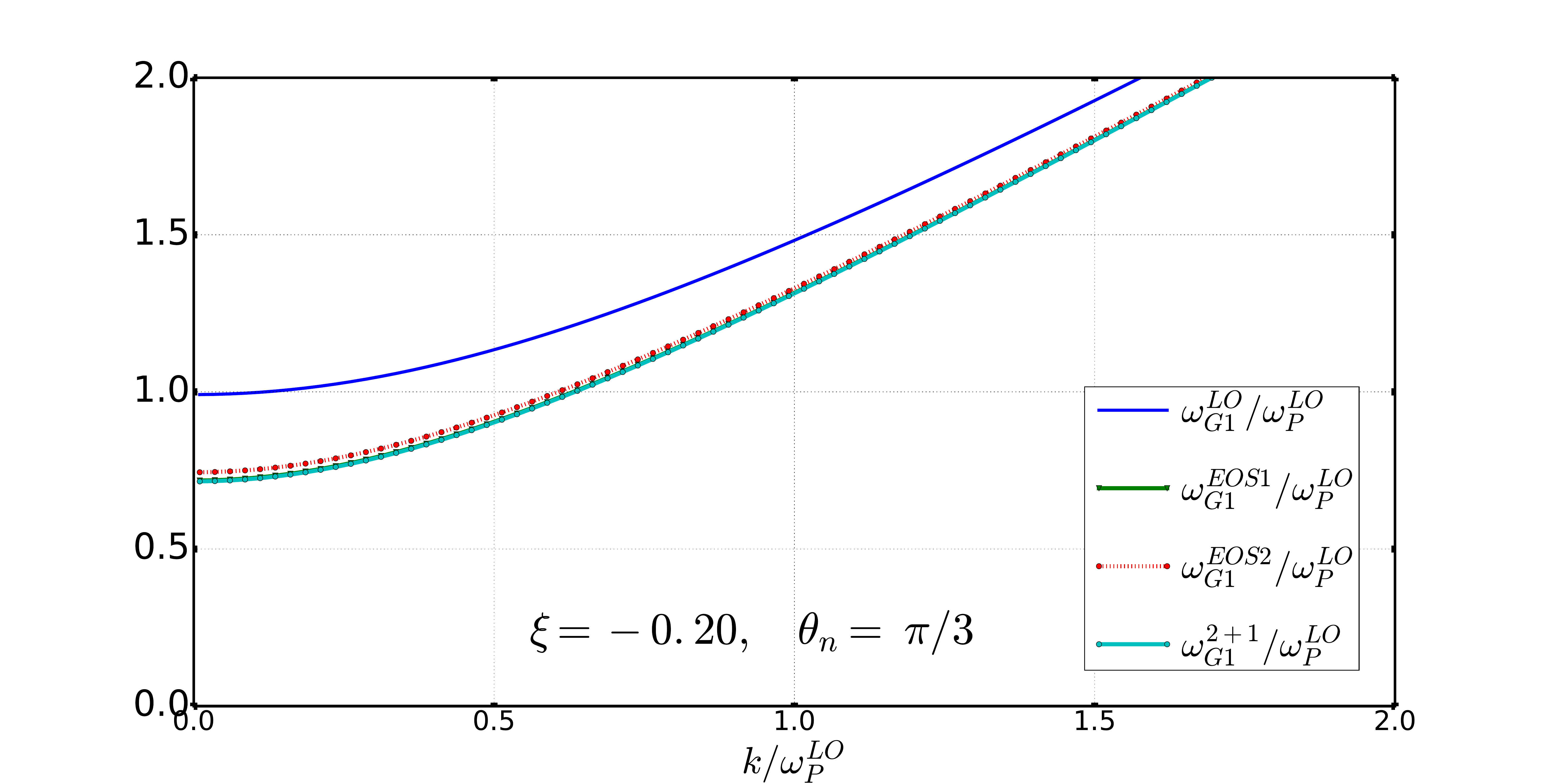}}
\subfloat{\includegraphics[height=5cm,width=6cm]{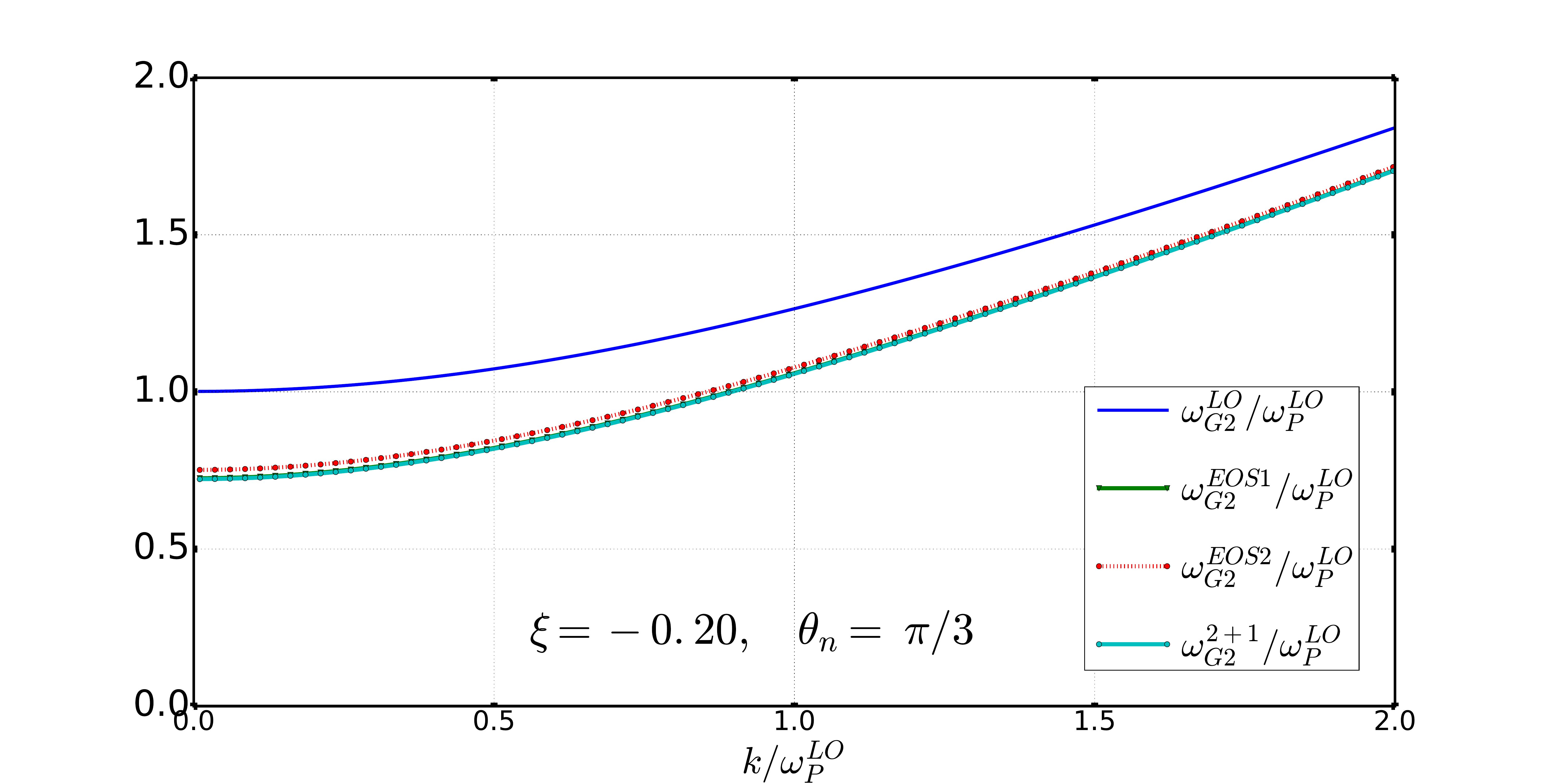}}
\caption{(color online) Dispersion curve of modes compared for various EOSs with different $\theta_n$ at fixed $\xi = -0.20$, $T_{c} = 0.17GeV$ and $T = 0.25GeV$.}
\label{fig:modes_N2}
 \end{figure*}
 
As discussed earlier, in order to find out the collective modes, we first have to obtain the 
dispersion equation by looking at the poles of the propagartor shown in Eq.(\ref{eq:propagator}). In order to do 
that we are first expressing,${\bf \Delta}^{-1}(k)$ in terms of structure functions, we have,
\begin{equation}
\begin{split}
{\bf \Delta}^{-1}(k) = (k^2 - \omega^2 + \alpha){\bf A} + (\beta - \omega^2){\bf B}\\ + \gamma {\bf C} + \delta {\bf  D} 
 \end{split}
\end{equation}
where {\bf A},{\bf B},{\bf C} and {\bf D} are the same 3-Tensors defined above.
Any symmetric 3-tensor ${\bf T}$ defined as:
\begin{equation}
{\bf T}=a\,{\bf A}+b\,{\bf B}+c\,{\bf C}+d\,{\bf D} 
\end{equation}
(a,b,c and d scalars)
will have the inverse as
\begin{equation}
{\bf T}^{-1}=a^{-1}{\bf A}+\frac{(a+c){\bf B}-a^{-1}(bc-\tilde n^2 k^2 d^2 ){\bf C}-d{\bf D}}{
b(a+c)- \tilde n^2 k^2 d^2 } \; .
 \end{equation}
Using the above formula, we obtain an expression for the propagator
\begin{equation}
\begin{split}
\Delta(k) =  \Delta_A  [{\bf A}-{\bf C}] + \Delta_G [(k^2 - \omega^2 + \\ \alpha + \gamma) {\bf B} +  (\beta-\omega^2) {\bf C}  - \delta {\bf D}]  
\end{split}
\end{equation}
with
\begin{equation}
\Delta_A^{-1}(k) = k^2 - \omega^2 + \alpha 
\label{mode_a}
\end{equation}
\begin{equation}
\Delta_G^{-1}(k) = (k^2 - \omega^2 + \alpha + \gamma)(\beta-\omega^2)-k^2 \tilde n^2 \delta^2 
\label{mode_g}
\end{equation}
As said earlier we can neglect $\delta^2$ in the linear order of approximation of $\xi$ and hence,
\begin{equation}
\Delta_G^{-1}(k) = (k^2 - \omega^2 + \alpha + \gamma)(\beta-\omega^2)
\end{equation}
we can split $\Delta_G^{-1}(k)$ into simpler form as,\\
\begin{equation}
\Delta_G^{-1}(k) = \Delta_{G1}^{-1}(k) ~\Delta_{G2}^{-1}(k)
\end{equation}
where
\ba
\Delta_{G1}^{-1}(k) = k^2 - \omega^2 + \alpha + \gamma,
\label{mode_g1}
\
~~~\Delta_{G2}^{-1}(k) = \beta-\omega^2
\label{mode_g2}
\ea

 In the next two subsections (A and B) we shall find the collective modes by solving the dispersion equations Eq.(\ref{mode_a}) and Eq.(\ref{mode_g2}). In the subsection C, we shall dicuss the results.

\subsection{A modes}
We refer A-modes when solving the dispersion equation(\ref{mode_a})as 
\begin{equation}
 \Delta_A^{-1}(k) = k^2 - \omega^2 + \alpha = 0. 
\end{equation}
In the limit $\omega^2 \gg k^2$ the equation is solved by
\ba
 \omega^2({\bf k})& =&\frac{m_D^{2}(T)}{3}(1 - \frac{\xi}{15}) + \frac{6}{5}\bigg[1 + \frac{\xi}{14}\Big(\frac{4}{15} + \cos^2\theta_n\Big)\bigg]k^2\nn&& + O\Big(k^4\Big)  
\ea
To look for pure imaginary solutions when $\omega^2 < k^2$ in equation(\ref{mode_a}), we substitute $\omega = i\Gamma$ with $\Gamma \in R$ and
assuming $\Gamma^2 \ll k^2$. With this approximation the $\alpha(\omega, {\bf k})$ becomes 
\ba
\alpha(\omega ,{\bf k})&=&-\frac{1}{3}\xi m_D^{2}(T)\cos^2\theta_n+\frac{\pi}{4}\bigg[1-\frac{\xi}{2}\Big(\frac{1}{3}\nn&&
-3 \cos^2\theta_n \Big)\bigg]m_D^2(T)\frac{|\Gamma|}{k} +O\Big(\frac{|\Gamma^2|}{k^2}\Big)
\ea 
and the solution  of the dispersion equation(\ref{mode_a}) is given as 
\ba
\Gamma_{A}({\bf k})&=& \pm\frac{1}{2}\bigg[\sqrt{\frac{\lambda_{A}^2}{k^2} + 4\Big(k^{2}_{A} - k^2\Big)} -\frac{\lambda_{A}}{k}\bigg]
\label{unstable_A}
\ea
where
\ba
k_{A} = m_D(T)|\cos\theta_n| Re\bigg(\sqrt{\frac{\xi}{3}}\bigg)
\ea
and
\begin{equation}
\lambda_{A} = \frac{\pi}{4}\bigg[1 - \frac{\xi}{2}\Big(\frac{1}{3} - 3\cos^2\theta_n \Big)\bigg]m_D^2(T) 
\end{equation}

\subsection{G modes}
We refer G-modes when solving the dispersion equation(\ref{mode_g})as 
\begin{equation}
\Delta_G^{-1}(k) = \Delta_{G1}^{-1}(k)~\Delta_{G2}^{-1}(k)=0
\end{equation}
which gives\\
\ba
\Delta_{G1}^{-1}(k)=k^2 - \omega^2 + \alpha + \gamma=0,~~~~\Delta_{G2}^{-1}(k)=\beta-\omega^2 = 0.\nn&& 
\ea
where $G1$ correspond to the transverse mode and $G2$ correspond to longitudinal mode.
In the limit $\omega^2\gg k^2$ the dispersion equation corresponding to $G1$ is solved by 

\ba
 \omega^2({\bf k})& =&\frac{m_D^{2}(T)}{3}\bigg[1 + \frac{\xi}{5}\Big(\frac{2}{3}
 -\cos^2\theta_n \Big)\bigg]\nn&& + \frac{6}{5}\bigg[1 - \frac{\xi}{5}\Big(\frac{23}{42}-\cos^2\theta_n \Big)\bigg]k^2+ O\Big(k^4\Big)  
\ea
For purely imaginary solution of $G1$ we again put $\omega = i\Gamma$ with $\Gamma \in R$ 
and assuming $\Gamma^2 \ll k^2$.The solution  of the dispersion equation
corresponding to $G1$ is obtained as,
\ba
\Gamma_{G1}({\bf k})&=& \pm\frac{1}{2}\bigg[\sqrt{\frac{\lambda_{G1}^2}{k^2} +
4\Big(k^2_{G1}- k^2\Big)}\nn&&  -\frac{\lambda_{G1}}{k}\bigg]
\label{unstable_G1}
\ea where
\ba
k_{G1} = m_D(T) Re\bigg(\sqrt{\frac{\xi}{3}(2\cos^2\theta_n-1)}\bigg)
\ea
and
\begin{equation}
\lambda_{G1} = \frac{\pi}{4}\bigg[1 - \frac{\xi}{2}\Big(\frac{7}{3} - 5\cos^2\theta_n \Big)\bigg]m_D^2(T)\end{equation}

In the same limit $\omega^2\gg k^2$ for $G2$, we have
\ba
 \omega^2({\bf k})& =&\frac{m_D^{2}(T)}{3}\bigg[1 + \frac{\xi}{5}\Big(-\frac{1}{3}
 +\cos^2\theta_n \Big)\bigg]\nn&& + \frac{3}{5}\bigg[1 + \frac{4\xi}{35}\Big(1-3\cos^2\theta_n \Big)\bigg]k^2+ O\Big(k^4\Big)  
\ea
which represents the longitudinal mode.
In the limit $\omega^2\ll k^2$, the imaginary solution of $G2$ does not exist. Since all the strucution functions are positive 
when putting $\omega = i\Gamma$, for real $\Gamma$ and  hence 
\ba
\Delta_{G2}^{-1}(k)=\beta-\omega^2 \neq 0.
\ea
So, we are having only two imaginary modes($\Gamma_A$ and $\Gamma_{G1}$) which are coming when $\omega^2\ll k^2$
 \begin{figure*}[]
  \centering
\subfloat{\includegraphics[height=7cm,width=9.4cm]{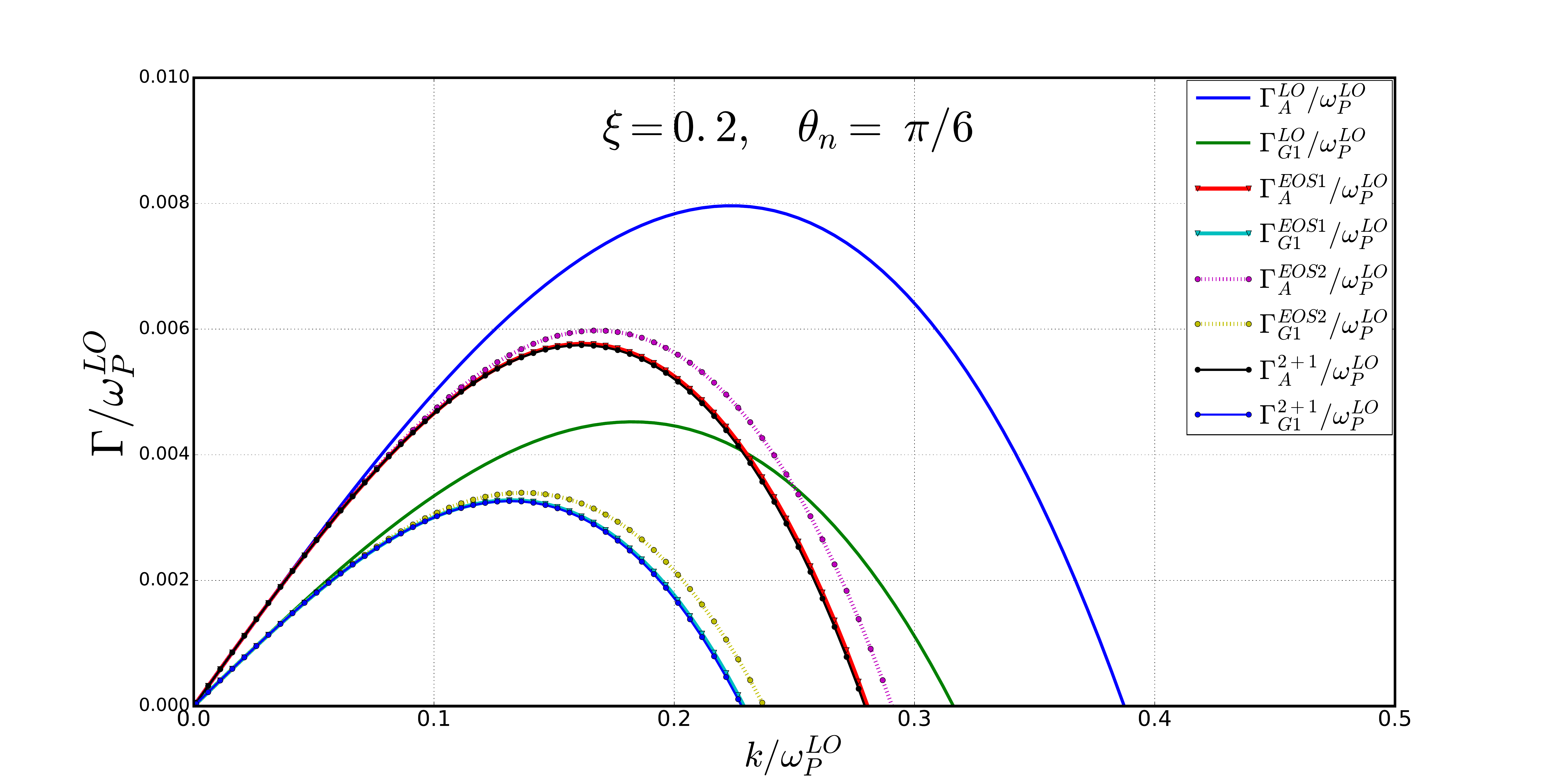}}
\subfloat{\includegraphics[height=7cm,width=9.4cm]{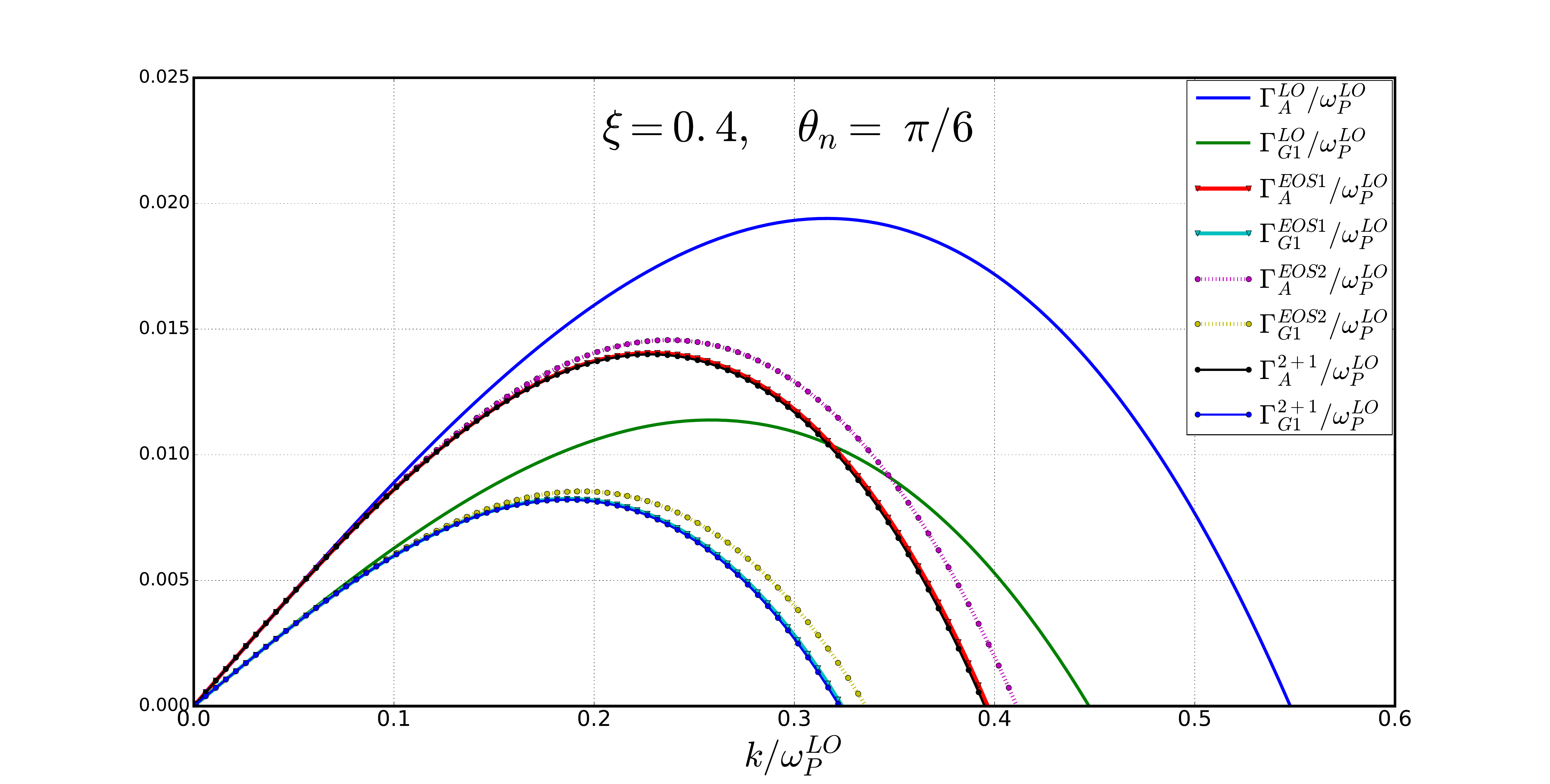}}
\caption{(color online)Dispersion curve of modes for various EOSs at fixed $T_{c} = 0.17GeV$ and $T = 0.25GeV$.}
\label{fig:im_modes_pi6}
\end{figure*}

 \begin{figure*}[]
  \centering
\subfloat{\includegraphics[height=7cm,width=9.4cm]{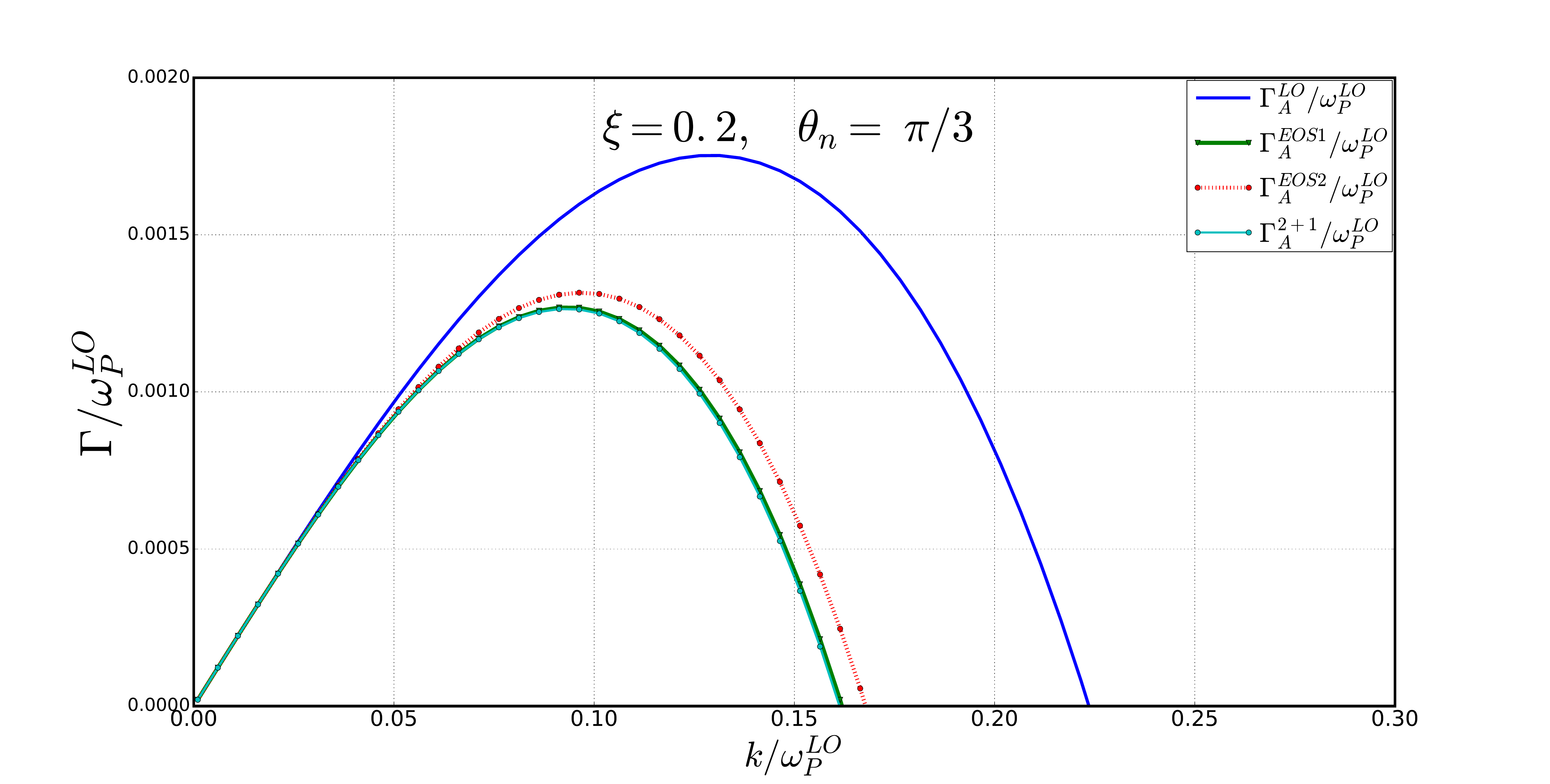}}
\subfloat{\includegraphics[height=7cm,width=9.4cm]{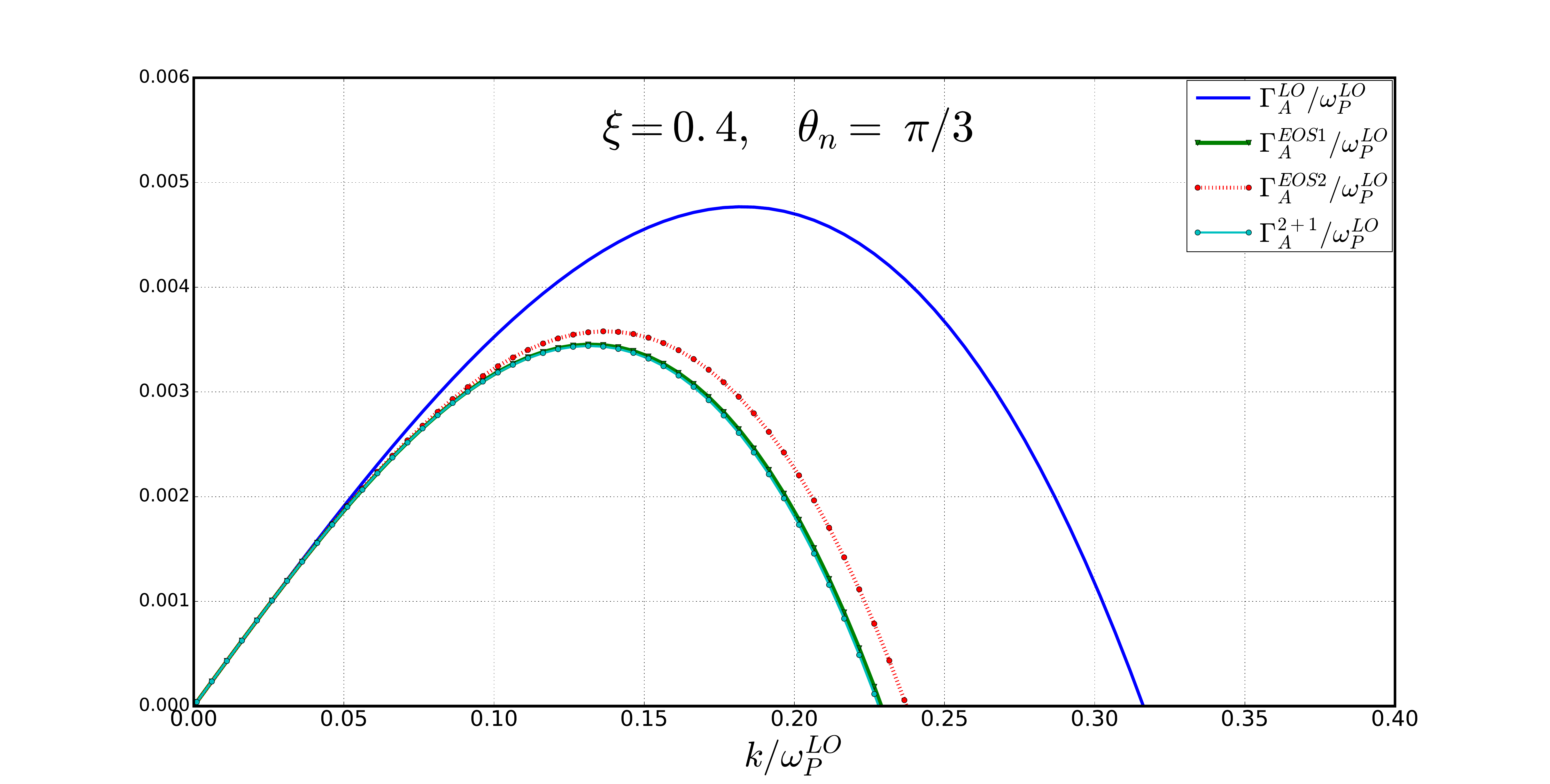}}
\caption{(color online)Dispersion curve of modes for various EOSs at fixed $T_{c} = 0.17GeV$ and $T = 0.25GeV$.}
\label{fig:im_modes_pi3}
\end{figure*}
 \begin{figure*}[]
  \centering
\subfloat{\includegraphics[height=7cm,width=9.4cm]{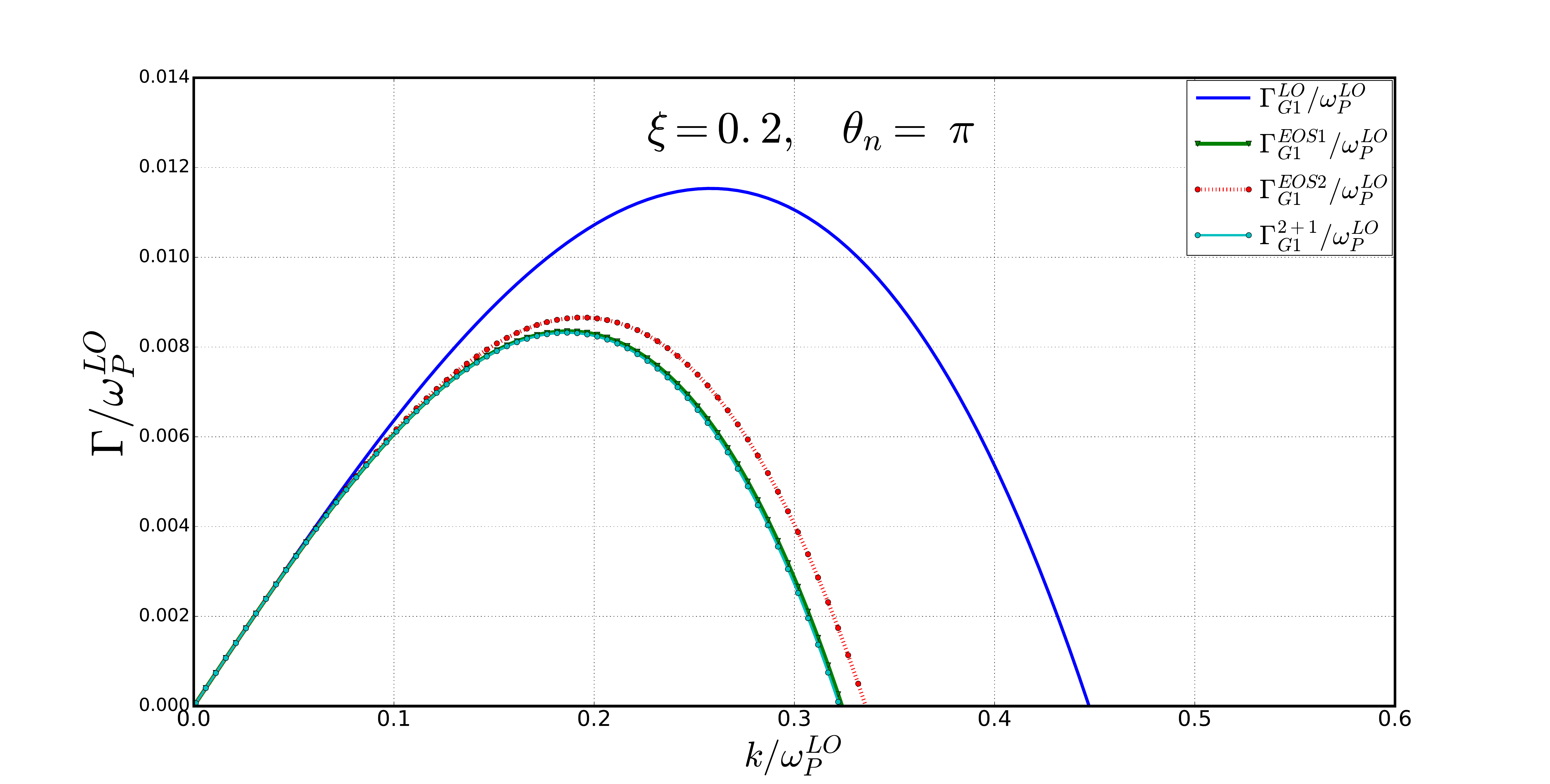}}
\subfloat{\includegraphics[height=7cm,width=9.4cm]{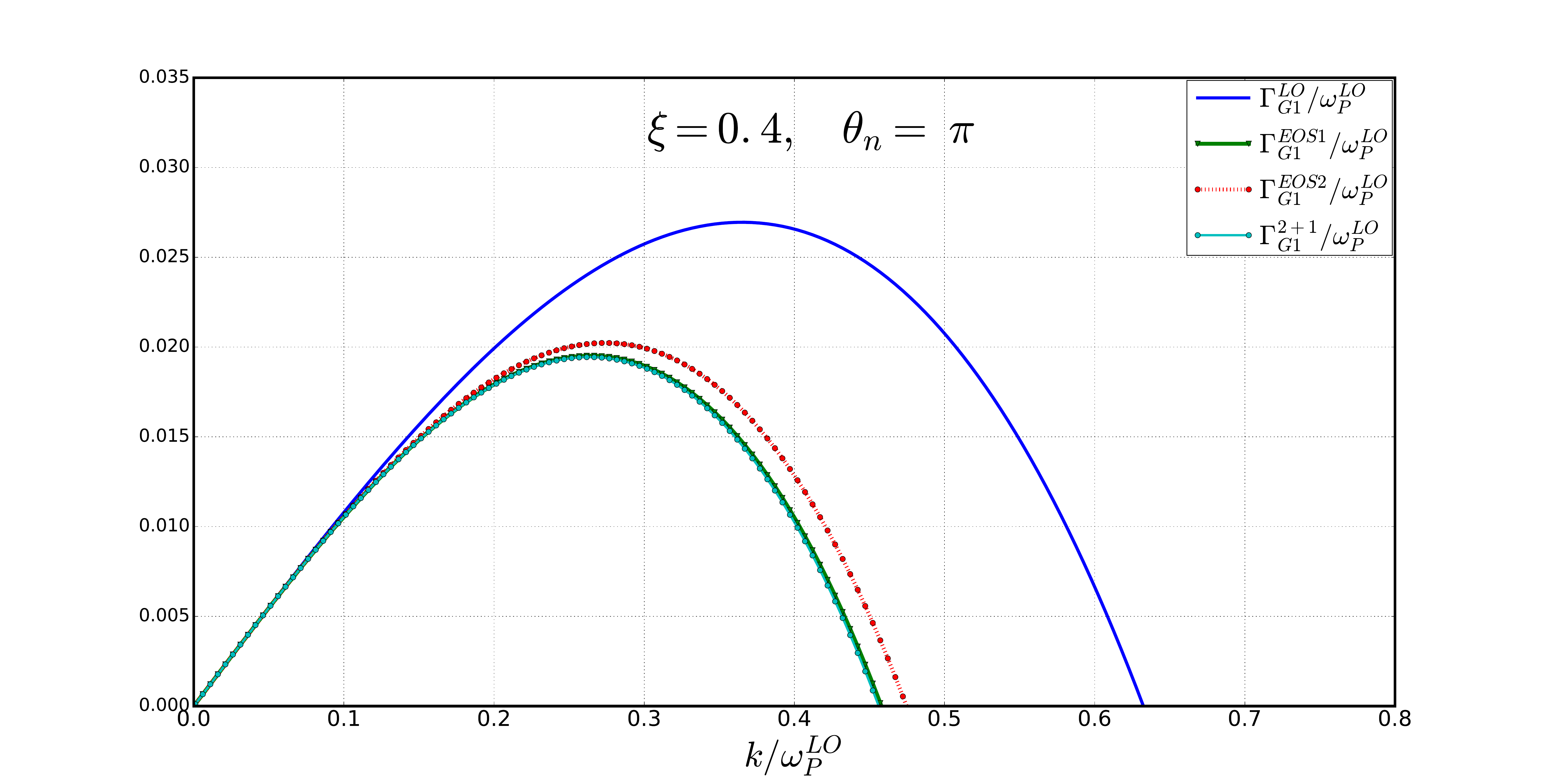}}
\caption{(color online)Dispersion curve of modes for various EOSs at fixed $T_{c} = 0.17GeV$ and $T = 0.25GeV$.}
\label{fig:im_modes_pi}
\end{figure*}
 \begin{figure*}[]
  \centering
\subfloat{\includegraphics[height=7cm,width=9.4cm]{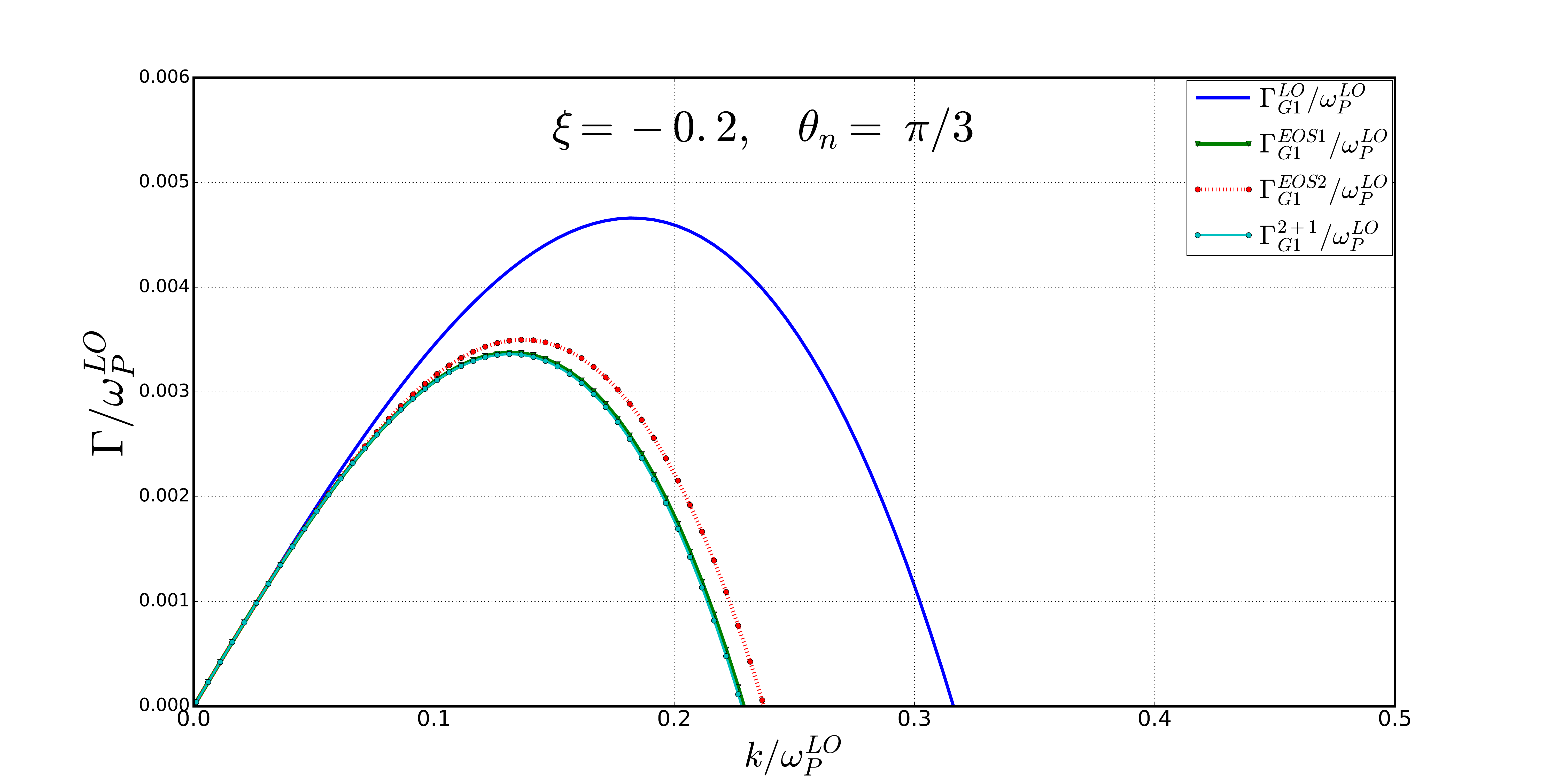}}
\subfloat{\includegraphics[height=7cm,width=9.4cm]{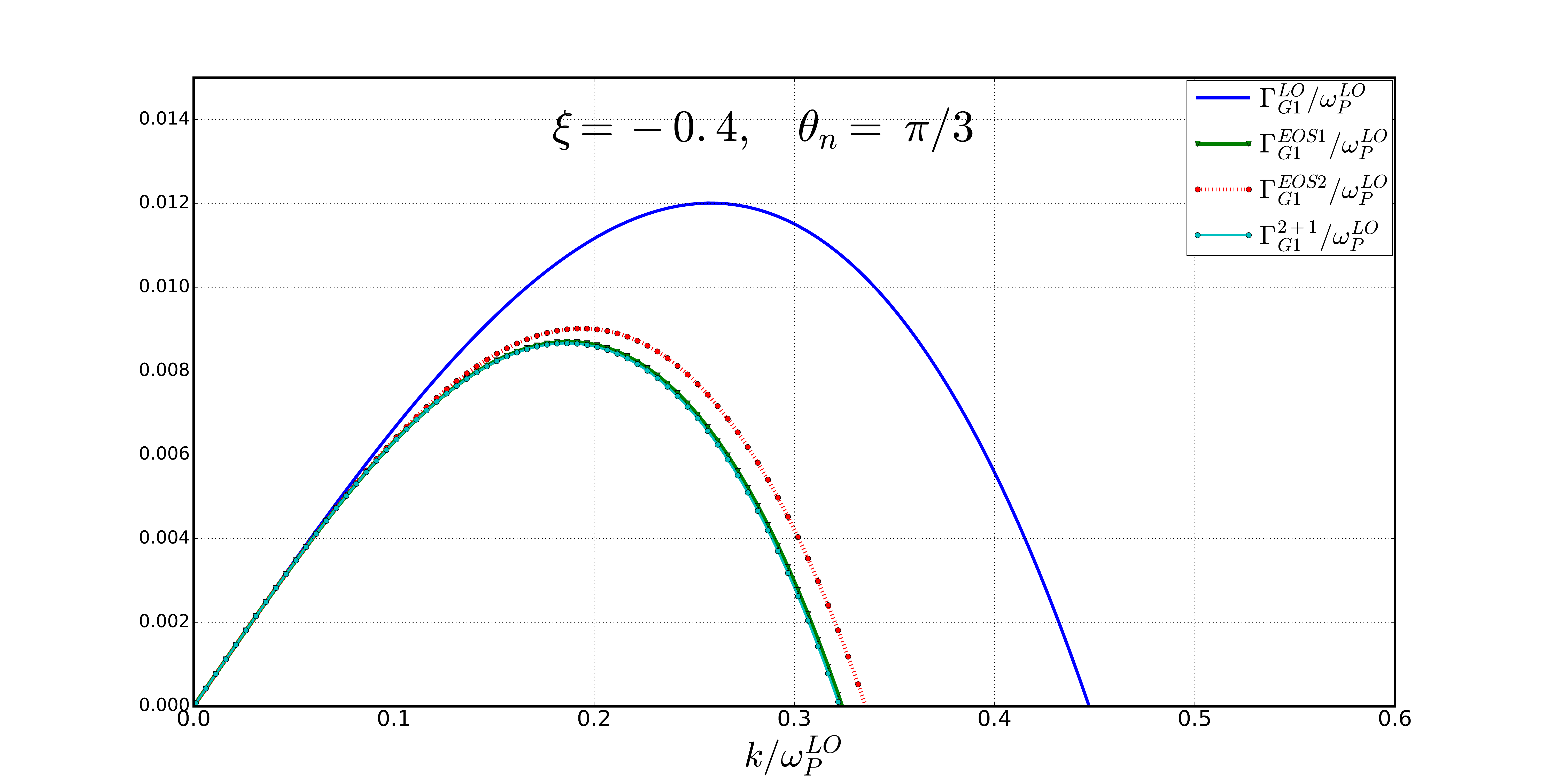}}
\caption{(color online)Dispersion curve of modes for various EOSs at fixed $T_{c} = 0.17GeV$ and $T = 0.25GeV$.}
\label{fig:im_modes_N_pi3}
\end{figure*}
 \begin{figure*}[]
  \centering
\subfloat{\includegraphics[height=7cm,width=9.4cm]{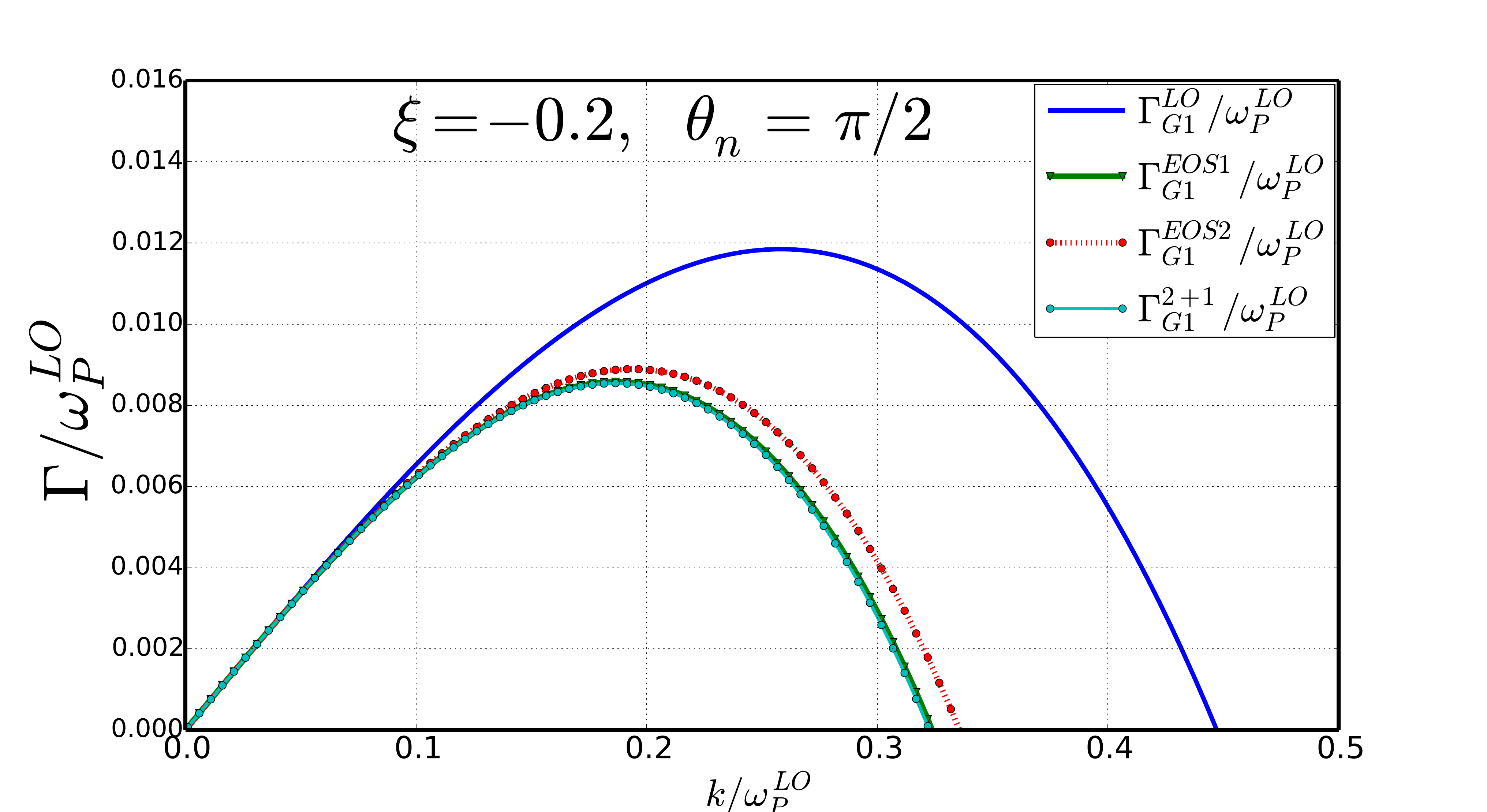}}
\subfloat{\includegraphics[height=7cm,width=9.4cm]{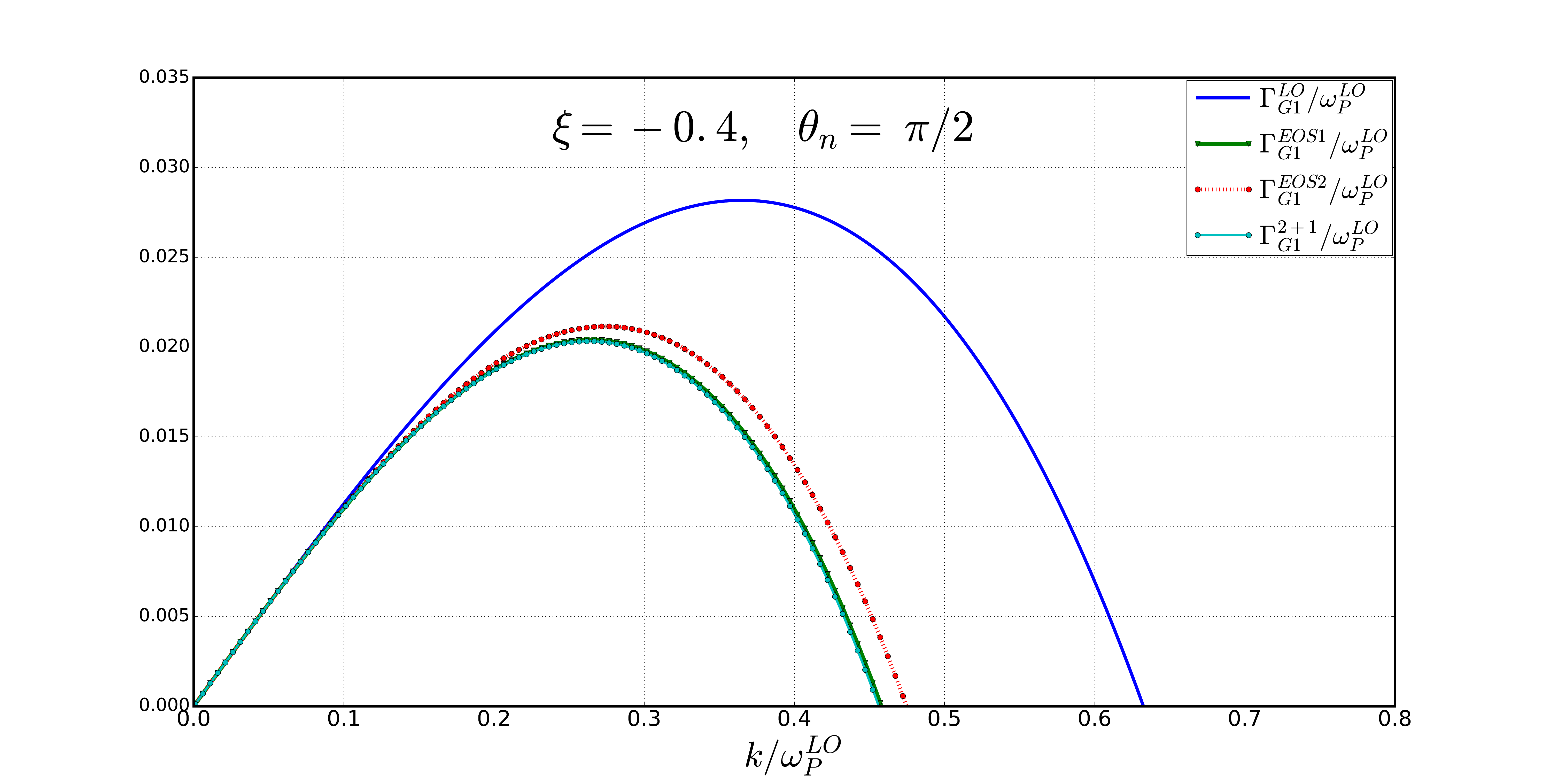}}
\caption{(color online)Dispersion curve of modes for various EOSs at fixed $T_{c} = 0.17GeV$ and $T = 0.25GeV$.}
\label{fig:im_modes_N_pi2}
\end{figure*}

\subsection{Results and Discussions}
In the small anisotropy limit, the dynamics of hot QCD medium in the transverse plane is understood by analysing the structure functions 
$\alpha$,  $\gamma$ and $\delta$. The response in the longitudinal directions is  captured in the structure function $\beta$. The forth structure function
 $\delta \sim  O(\xi)$, enters in the dispersion equation as $\delta^2$. Therefore, it has been ignored henceforth. Next, to highlight the effects of various EOSs, we consider these quantities 
 scaled with the plasma frequency,  $\omega_p^{LO}\equiv m_D^{LO}/\sqrt{3}$ as:
 \ba
 \tilde{\alpha}=\frac{\alpha}{(\omega_p^{LO})^2}, \  \ \tilde{\beta}=\frac{\beta}{(\omega_p^{LO})^2},\  \  \tilde{\gamma}=\frac{\gamma}{(\omega_p^{LO})^2}.
 \ea
 The $\omega/k$ dependence of $\tilde{\alpha}$ for isotropic($\xi = 0.0$), $\xi = 0.2$ and $\xi = -0.2$ is shown in 
Fig.~\ref{fig:alpha}  at $\theta_n =\pi/3$, $T_{c} = 0.17GeV$ and $T = 0.25 GeV$. Similarly,  $\tilde{\beta}$  is shown in Fig.~\ref{fig:beta} and $\tilde{\gamma}$ is shown 
in Fig.~\ref{fig:gamma}. The leading order HTL results are shown in the same figures for the purpose of comparison. It is to be noted that the 
realistic EOSs have modified the results quite significantly for both real and imaginary parts of all the three structure functions . The EOS1, EOS2 and LEOS  results are quite closer while, we look them in contrast to LO case.  Notably, for real $\omega$ when $\omega < k$,  all  the three  structure functions ($\tilde{\alpha}$, $\tilde{\beta}$ and $\tilde{\gamma}$) are complex while  for $\omega > k$,  they only possess non-vanishing real parts. 

The dispersion curves for various collective modes have been shown in Figs.~\ref{fig:modes_0}-\ref{fig:modes_N2} for real $\omega$ and $\omega> k$ 
at  $\xi = 0 , 0.2,  -0.2$  and  $\theta_n = 0 , \pi/3$.  We have chosen, $T_{c} = 0.17GeV$ and $T = 0.25GeV$ throughout.
To visualise the EOS effects in contrast to LO results, the modes have been scaled with $\omega_p^{LO}$.
Fig.~\ref{fig:modes_0} is plotted for the isotropic case, 
the dispersion curves of collective modes($\omega_{A}$, $\omega_{G1}$, $\omega_{G2}$)  have been shown 
for two different values of the angles ($\theta_n=0, \pi/3$).
The spectrum of weakly squeezed plasma, $\xi = 0.2$, have been plotted in Fig.\ref{fig:modes_2}.
Similarly, weakly  stretched case, $\xi = -0.2$, is shown in Fig.~\ref{fig:modes_N2}. In all the three cases,
 we found similar pattern but the numbers are visibly different in the case of different EOSs in contrast to the LO results.
Notably, the $G2$ modes in both isotropic and anisotropic case croses the light cone($\omega = k$). This effect is also been mentioned in~\cite{Carrington:2014bla}. 
For non-zero $\xi$, the propagator also has poles along the imaginary $\omega$ axis when $\omega^2\ll k^2$.
We have found the dispersion relation shown in Eq.(\ref{unstable_A}) and Eq.(\ref{unstable_G1}) by taking 
$\omega = i\Gamma$ with $\Gamma$ real valued and plotted $\Gamma/\omega^{LO}_p$ with respect to $k/\omega^{LO}_p$.

We have plotted the spectrum of weakly squeezed plasma, from
Fig.\ref{fig:im_modes_pi6} to Fig.\ref{fig:im_modes_pi} 
 at different angle $\theta_n$. 
In each plot we got the results in different range. Again in order to highlight the effects of 
various EOSs, in all the cases, along with the LO results, we have also plotted the 
results coming out using EOS1, EOS2 and LEOS.
Fig.\ref{fig:im_modes_pi6} is
plotted for $\xi =0.2$ and $\xi = 0.4$, respectively in the left and right panel, at $\theta_n = \pi/6$. 
As one can easily see that in Fig.\ref{fig:im_modes_pi6}, we got both the modes({\it i.e.,} 
$\Gamma_A$ and $\Gamma_{G1}$). Same in Fig.\ref{fig:im_modes_pi} also for $\theta_n = \pi$.
But as shown in Fig.\ref{fig:im_modes_pi3}, which is plotted for $\theta_n = \pi/3$,
we got only $\Gamma_A$ mode.

Weakly stretched spectrum of hot QCD plasma is plotted in Fig.\ref{fig:im_modes_N_pi3} and
Fig.\ref{fig:im_modes_N_pi2} for $\theta_n = \pi/3$ and $\theta_n = \pi/2$ respectively.
In this case($\xi$ negative), we got only $\Gamma_{G1}$ mode while the $\Gamma_{A}$ mode was absent.
The 
Fig.\ref{fig:im_modes_N_pi3}(Left panel) and Fig.\ref{fig:im_modes_N_pi3}(Right panel) are plotted for $\xi = -0.2$
and $\xi = -0.4$, respectively. In the same way Fig.\ref{fig:im_modes_N_pi2}(Left panel) and 
Fig.\ref{fig:im_modes_N_pi2}(Right panel) are plotted for $\xi = -0.2$ and $\xi = -0.4$, respectively.
In total,  three  real modes for $\omega^2 \gg k^2$ and at-most two imaginary modes for $\omega^2 \ll k^2$ are seen to be there. All of them 
are  seen to be significantly influenced by the realistic EOS as compared to the LO/Ideal EOS. Moreover, the imaginary modes have seen to exist only 
for certain values of $\theta_n$ similar to what has already been observed in Refs. \cite{Romatschke:2003ms,Carrington:2014bla}.

\section{Summary and Conclusions}
In conclusion, the gluon polarization tensor  for  a hot QCD/QGP medium is obtained 
within linear transport theory. The dispersion equations and gluon collective modes are obtained in terms of the polarization tensor 
while invoking the classical Yang-Mills equation of motion. The hot QCD medium  interactions are modelled by exploiting the quasi-particle
 description of the hot QCD equations of state in terms of quark and gluon effective fugacities contained by their respective effective
 momentum distributions. The EOSs under considerations include both improved perturbative QCD and  (2+1)-flavor lattice QCD with
 physical quark-masses. This enables one to distinguish different EOSs in RHIC in the context of collective behaviour of hot QCD plasma. 
The collective mode analysis of the hot QCD plasma revealed that interactions induce significant modifications to  the modes 
(temperature dependence). However, the $\omega-k$ dependence of the longitudinal and transverse gluon polarization remains intact (same as that while invoking an ideal EOS).
An extension to the anisotropic hot QCD medium is done by rescaling the above mentioned isotropic quasi-particle distributions. The effects of hot QCD medium along with the anisotropy to the 
collective modes are studied in detail. The hot QCD medium as well as presence of anisotropy are seen to have significant impact the collective plasma properties of the medium.

An immediate future extension of the work would be to include the collectivity and near perfect fluidity of the hot QCD medium/QGP and study its impact on collective plasma properties while analyzing the 
quark-antiquark effective modes. The realization of hot QCD medium in terms of a refractive index would also be taken up in future. Equally importantly, obtaining 
an expression for the inter-quark potential in this medium and its phenomenological aspect will be another direction where the investigations focus will on.

\section{Acknowledgements}
 VC would like to sincerely acknowledge DST, Govt. of India for Inspire Faculty Award -IFA13/PH-15
 and Early Carrier Research Award(ECRA/2016) Grant. SM would like to acknowledge IIT Gandhinagar for Institite 
postdoctoral fellowoship and research infrastructure. We would like to acknowledge people of INDIA for their generous
 support for the research in fundamental sciences in the country.

{}

\begin{thebibliography}{99}

\bibitem{expt_rhic}
J. Adams {\it et al.}  (STAR Collaboration), Nucl.  Phys.  A {\bf 757}, 102 (2005);
 K. Adcox {\it et al.} PHENIX Collaboration, Nucl. Phys.  A {\bf 757}, 184 (2005);
 B.B. Back {\it et al.} PHOBOS Collaboration, Nucl. Phys. A  {\bf 757}, 28 (2005);
 I. Arsene {\it et al.} BRAHMS Collaboration, Nucl. Phys. A {\bf 757}, 1 (2005). 

\bibitem{expt_lhc}
K. Aamodt {\it et al.} (The Alice Collaboration), Phys. Rev.
Lett. {\bf 105}, 252302 (2010);
 Phys. Rev.  Lett. {\bf 105}, 252301 (2010); Phys. Rev. Lett. {\bf 106}, 032301
(2011).


  
\bibitem{Ryu}
S.  Ryu, J. -F.  Paquet, C.  Shen, G. S. Denicol, B. Schenke, S. Jeon,  C. Gale, Phys.  Rev.   Lett.  {\bf 115}, 132301 (2015).

\bibitem{Denicol1}
G.  Denicol, A.  Monnai, B.  Schenke,   Phys. Rev.  Lett. {\bf 116}, 212301 (2016).

\bibitem{Chu:1988wh}
  M.~C.~Chu and T.~Matsui,
  Phys.\ Rev.\ D {\bf 39} (1989) 1892.

\bibitem{Landau:1984}{L.D.Landau and E.M.Lifshitz}(1984),
  {Electrodynamics of continuous media}
  ({Butterworth-Heinemann}, (1984).

\bibitem{Bellac:1996}{Bellac}(1996)
M.~L.Bellac,{Cambridge University Press},(1996).

\bibitem{Koike:1991mf}
  Y.~Koike,
  AIP Conf.\ Proc.\  {\bf 243} (1992) 916.


 \bibitem{Mrowczynski:1993qm}
  S.~Mrowczynski,
  Phys.\ Lett.\ B {\bf 314} (1993) 118.

 \bibitem{Mrowczynski:1994xv}
  S.~Mrowczynski,
  Phys.\ Rev.\ C {\bf 49} (1994) 2191.

  \bibitem{Mrowczynski:1996vh}
  S.~Mrowczynski,
  Phys.\ Lett.\ B {\bf 393} (1997) 26.

  
\bibitem{dm_rev1}
 Daniel F. Litim, C. Manual, Phys. Rep.  {\bf 364}, 451 (2002).

 \bibitem{dm_rev2} 
  J. P. Blaizot, E. Iancu, Phys. Rep. {\bf 359}, 355 (2002).
 
\bibitem{Mrowczynski:2000ed}
  S.~Mrowczynski and M.~H.~Thoma,
  Phys.\ Rev.\ D {\bf 62} (2000) 036011.

  \bibitem{Mrowczynski:2004kv}
  S.~Mrowczynski, A.~Rebhan and M.~Strickland,
  Phys.\ Rev.\ D {\bf 70} (2004) 025004.
  
\bibitem{Weibel:1959zz}
  E.~S.~Weibel,
  Phys.\ Rev.\ Lett.\  {\bf 2} (1959) 83.
  
  \bibitem{Romatschke:2003ms}{Romatschke and Strickland}(2003)
{Phys. Rev. D} \textbf{68}, (2003).
  
  
 
 \bibitem{Romatschke:2004jh}{Romatschke and Strickland}(2004)
{Phys. Rev. D} \textbf{70}, (2004).
  
  \bibitem{Mrowczynski:2005ki}
  S.~Mrowczynski,
  Acta Phys.\ Polon.\ B {\bf 37} (2006) 427.
  
  
 \bibitem{Arnold:2003rq}
  P.~B.~Arnold, J.~Lenaghan and G.~D.~Moore,
  JHEP {\bf 0308} (2003) 002.
  
   
 
  \bibitem{Attems:2012js}
  M.~Attems, A.~Rebhan and M.~Strickland,
  Phys.\ Rev.\ D {\bf 87} (2013) no.2,  025010.
  
\bibitem{Florkowski:2012as} 
  W.~Florkowski, R.~Maj, R.~Ryblewski and M.~Strickland,
  Phys.\ Rev.\ C {\bf 87}, no. 3, 034914 (2013).
  
 \bibitem{Dumitru:2007hy}
  A.~Dumitru, Y.~Guo and M.~Strickland,
  Phys.\ Lett.\ B {\bf 662} (2008) 37.
  
\bibitem{Martinez:2008di}
  M.~Martinez and M.~Strickland,
  Phys.\ Rev.\ C {\bf 78} (2008) 034917.



  
\bibitem{Schenke:2006yp}
  B.~Schenke and M.~Strickland,
  Phys.\ Rev.\ D {\bf 76} (2007) 025023.
  
\bibitem{Kobes:1990dc}
  R.~Kobes, G.~Kunstatter and A.~Rebhan,
  Nucl.\ Phys.\ B {\bf 355} (1991) 1.
   
  \bibitem{chandra_quasi1}
  Vinod Chandra, R. Kumar, V. Ravishankar, Phys. Rev.  C {\bf 76}  (2007) 054909, [Erratum: Phys. Rev.  C {\bf 76}, 069904  (2007).
  Vinod Chandra, A. Ranjan, V. Ravishankar,  Eur. Phys. J.  A {\bf 40}, 109-117  (2009).
    
  \bibitem{chandra_quasi2} Vinod Chandra, V. Ravishankar  
  Phys. Rev.  D {\bf 84}, 074013   (2011). 


\bibitem{Carrington:2014bla}
  M.~E.~Carrington, K.~Deja and S.~Mrowczynski,
  Phys.\ Rev.\ C {\bf 90} (2014) no.3,  034913.

 
\bibitem{Mitra:2016zdw} 
  S.~Mitra and V.~Chandra,
  Phys.\ Rev.\ D {\bf 94}, no. 3, 034025 (2016).
  
\bibitem{Agotiya:2016bqr} 
  V.~K.~Agotiya, V.~Chandra, M.~Y.~Jamal and I.~Nilima,
  Phys.\ Rev.\ D {\bf 94}, no. 9, 094006 (2016)
  

\bibitem{Chandra:2010xg} 
  V.~Chandra and V.~Ravishankar,
  Nucl.\ Phys.\ A {\bf 848}, 330 (2010).


\bibitem{Chandra:2015gma} 
  V.~Chandra and S.~K.~Das,
  Phys.\ Rev.\ D {\bf 93}, no. 9, 094036 (2016).

\bibitem{Das:2012ck} 
  S.~K.~Das, V.~Chandra and J.~e.~Alam,
  J.\ Phys.\ G {\bf 41}, 015102 (2013).


\bibitem{Chandra:2015rdz} 
  V.~Chandra and V.~Sreekanth,
  Phys.\ Rev.\ D {\bf 92}, no. 9, 094027 (2015).

\bibitem{Chandra:2016dwy} 
  V.~Chandra and V.~Sreekanth,
  arXiv:1602.07142 [nucl-th].

  \bibitem{polya}
 M. D\'Elia, A. Di Giacomo, E. Meggiolaro, Phys. Lett.  {\bf B 408}, 315 (1997); Phys. Rev. D {\bf  67}, 114504 (2003); P. Castorina, M. Mannarelli, Phys. Rev. C {\bf 75}, 054901 (2007);  Phys.  Lett.  {\bf B 664}, 336 (2007)., 
 Paolo Alba {\it et al.}, Nucl. Phys.  A {\bf 934}, 41-51 (2014).  

\bibitem{pnjl}
A. Dumitru, R. D. Pisarski, Phys. Lett.  {\bf B 525}, 95 (2002);
K. Fukushima, Phys.  Lett.  {\bf B 591}, 277 (2004); 
S. K. Ghosh {\it et. al}, Phys.  Rev.  D {\bf 73}, 114007 (2006);
H. Abuki, K. Fukushima, Phys. Lett.  {\bf B 676}, 57 (2006); 
H. M. Tsai, B. M\"{u}ller,  J. Phys.  G  {\bf 36}, 075101 (2009).

  
  \bibitem{effmass1}
 A. Peshier  {\it et. al}, Phys.  Lett.  {\bf B 337}, 235 (1994); Phys.
Rev. D {\bf 54}, 2399 (1996).
 
 \bibitem{effmass2} A. Peshier, B. K\"{a}mpfer, G. Soff, Phys. Rev. C {\bf 61},
045203 (2000); Phys. Rev.  D {\bf 66}, 094003 (2002); V. M. Bannur, Phys. Rev. C {\bf 75}, 044905 (2007); {\it ibid}. C {\bf 78}, 045206 (2008); JHEP {\bf 0709}, 046 (2007); A. Rebhan, P. Romatschke, Phys. Rev. D {\bf 68}, 0250022 (2003); 
 M. A. Thaler, R. A. Scheider, W. Weise, Phys. Rev. C {\bf 69}, 035210 (2004);  K. K. Szabo, A. I. Toth, JHEP  {\bf 0306}, 008 (2003). 


\bibitem{Rozynek:2016ykp} 
  J.~Rozynek and G.~Wilk,
  Eur.\ Phys.\ J.\ A {\bf 52}, no. 9, 294 (2016)
  
\bibitem{Bluhm}
  M.~Bluhm, B.~Kampfer and K.~Redlich,
  Phys.\ Rev.\ C {\bf 84} (2011) 025201.
 
 \bibitem{chandra_eta} 
 Vinod Chandra, V. Ravishankar, Eur.  Phys.  J. C  {\bf 64}, 63-72  (2009) ; {\it ibid.} C {\bf 59}, 705-714  (2009).
  
\bibitem{chandra_etazeta}
Vinod Chandra, Phys. Rev. D {\bf 86} 114008 (2012); {\it ibid.}, D {\bf 84}, 094025  (2011).
 
\bibitem{PJI}
P. Chakraborty, J. I.  Kapusta , 
Phys.  Rev.  C  {\bf 83},  014906 (2011).

 \bibitem{Mkap} 
 Albright and J. I. Kapusta,  Phys. Rev. C {\bf 93}, 014903 (2016).


\bibitem{Greco} A. Puglisi, , S. Plumari,  V. Greco, Phys. Lett.  B {\bf 751}, 326-330 (2015).



  \bibitem{cheng}
  M. Cheng {\bf et. al}, Phys. Rev. D {\bf 77}, 014511  (2008). 
  
  \bibitem{leos1_lat}

 A. Bazavov {\it et al.}, Phys. Rev. D {\bf 80}, 014504 (2009); 
M. Cheng {\it et al.}, Phys. Rev. D {\bf 81}, 054504 (2010); 
S. Borsanyi {\it et al.} , J. High Energy Phys. {\bf 11} (2010) 077; Y.
Aoki {\it et al.}, J. High Energy Phys. {\bf 01} (2006) 089; J. High
Energy Phys. {\bf 06}  (2009) 088; 
S. Borsanyi {\it et al.}, J. High Energy Phys. {\bf 09}  (2010) 073;
A.  Bazavov {\it et al.},   Phys.  Rev.  D {\bf 90}, 094503 (2014);
S.  Borsanyi, Z.  Fodor, C. Hoelbling, S. D. Katz, S. Krieg, K. K. Szabo,   Phys.  Lett.  {\bf B 370}, 99-104,  (2014).
\bibitem{kaj}
K. Kajantie, M. Laine, K. Rummukainen and Y. Schroder,
Phys. Rev. D {\bf 67},  105008 (2003).  

 \bibitem{qcd_coupling}  M. Laine and Y. Sch\"{o}der,  J HEP {\bf 0503} (2005) 067.
 
 \bibitem{zhai} 
P.  Arnold and Chengxing Zhai, Phys.  Rev.  D {\bf  50},  7603;(1994);
{\it ibid} {\bf 51},   1906( 1995).

\bibitem{kastening}
Chengxing Zhai and B. Kastening, Phys. Rev. D {\bf 52}, 7232. (1995).

\bibitem{dmass1}
Kelly  {\it et. al}, Phys. Rev. Lett.  {\bf 72}, 3461 (1994), Phys. Rev. D {\bf 50}, 4209 (1995).

  \bibitem{Elze:1989un}
  H.~T.~Elze and U.~W.~Heinz,
  Phys.\ Rept.\  {\bf 183} (1989) 81.
  
 
\bibitem{Liu:2011if} 
  J.~Liu, M.~j.~Luo, Q.~Wang and H.~j.~Xu,
  Phys.\ Rev.\ D {\bf 84}, 125027 (2011).

 

 
 

 
 \end{thebibliography}
\end{document}